\documentclass[prd,twocolumn,showpacs,floatfix,nofootinbib]{revtex4}
\usepackage{graphicx,color,dcolumn,booktabs,bm}
\usepackage{longtable,lscape}
\usepackage{txfonts}
\usepackage{overpic}
\usepackage{amssymb}
\usepackage{indentfirst}

\begin{document}

\title{Photoproduction of the charged charmoniumlike $Z_{c}^{+}(4200)$}
\author{Xiao-Yun Wang$^{1,2,3}$}
\thanks{xywang@impcas.ac.cn}
\author{Xu-Rong Chen$^{1,3}$}
\affiliation{$^1$Institute of Modern Physics, Chinese Academy of Sciences, Lanzhou
730000, China\\
$^2$University of Chinese Academy of Sciences, Beijing 100049, China\\
$^3$Research Center for Hadron and CSR Physics, Institute of Modern Physics
of CAS and Lanzhou University, Lanzhou 730000, China}
\author{Alexey Guskov}
\thanks{avg@jinr.ru}
\affiliation{$^1$Joint Institute for Nuclear Research, Dubna 141980, Russia}

\begin{abstract}
In this work, inspired by the observation of charmoniumlike $Z_{c}^{+}(4200)$%
, we study the photoproduction of charged charmoniumlike $Z_{c}^{+}(4200)$
with an effective Lagrangian approach and the Regge trajectories model. The
numerical results indicate that the Reggeized treatment can lead to a lower
total cross section of the $Z_{c}^{+}(4200)$ photoproduction and the peak
position of cross section was moved to the higher energy point when the
Reggeized treatment was added. {Moreover, using the data from the COMPASS
experiment and presented theoretical predictions, an upper limit of the
decay width of $Z_{c}(4200)\rightarrow J/\psi \pi $ is estimated.} The
relevant results not only shed light on the further experiment of searching
for the charmoniumlike $Z_{c}(4200)$ state via meson photoproduction, but
also provide valuable informations for having a better comprehension of the
nature of charmoniumlike $Z_{c}(4200)$ state.
\end{abstract}

\maketitle

\section{Introduction}

As of now, most of hadrons can be well described by the classical
constituent quark model in the picture of $q\bar{q}$ for mesons and $qqq$
for baryons. However, according to the quantum chromodynamics, the exotic
states (such as the multiquark states, molecule states etc.) are also
allowed to exist in our Universe. Therefore, searching and explaining these
exotic states arouse great interest among researchers.

In the experiments, a series of charmoniumlike and bottomoniumlike states
referred to $XYZ$ have been observed \cite%
{belle03,lhc13,ba05,belle07,bes13,belle110,bes14,bes111,belle08,belleprd,belle100,lhc14,belle14,ab12,ia12,iab}%
. Especially, those charged $Z$ states are even more exotic since they have
a minimal quark content of $\left\vert c\bar{c}u\bar{d}\right\rangle $ ($%
Z_{c}^{+}$) or $\left\vert b\bar{b}u\bar{d}\right\rangle $ ($Z_{b}^{+}$)
\cite{lx14,ae14,sl14,mn10,mn14}. Later, some neutral $Z$ states (including $%
Z_{c}^{0}(3900)$, $Z_{b}^{0}(10610)$ and $Z_{c}^{0}(4020)$) were reported by
experiments \cite{tx13,zb13,zc14}, which provide important informations of
confirming and understanding the exotic $Z$ states.

On theoretical aspects, these exotic states are interpreted as a hadronic
molecule, a tetraquark, hadrocharmonium or just a cusp effect. \cite%
{lx14,ae14,sl14,mn10,mn14,zsl05,zsl10,ln12,hh06,de06,nb06,qw13,eb13,dy13,cf14,fa14,iv12,db11,rdm07,lm14,zl14,wei11,wei12,wei15}%
, \textit{et al. }Moreover, several hidden charm baryons composed by $%
\left\vert c\bar{c}qqq\right\rangle $ have been predicted and investigated
\cite{wjj10,wjj11,xyw15,xy2015}. These studies enriched the picture of
exotic states.

Recently, Belle Collaboration claimed that a new charged charmoniumlike $%
Z_{c}^{+}(4200)$ was observed in the invariant mass spectrum of $J/\psi \pi
^{+}$ with a significance of 6.2$\sigma $ \cite{belle14}. Its mass and width
are $M_{Z_{c}(4200)}=4196_{-29-13}^{+31+17}$ MeV/$c^{2}$ and $\Gamma
_{Z_{c}(4200)}=370_{-70-132}^{+70+70}$ MeV \cite{belle14}, respectively.
Meanwhile, the quantum number of $Z_{c}^{+}(4200)$ was determined to be $%
J^{P}=1^{+}$ since other hypotheses with $J^{P}\in \left\{
0^{-},1^{-},2^{-},2^{+}\right\} $ were excluded \cite{belle14}. In Ref. \cite%
{zl14}, the calculations show that $Z_{c}(4200)$ is a strong candidate of
the lowest axial-vector tetraquark state within the framework of the
color-magnetic interaction. In Refs. \cite{wei11,wei12,wei15}, using the QCD
sum rule approach, the relevant results also support the tetraquark
interpretation of $Z_{c}(4200)$. Besides, the $Z_{c}(4200)$ was described as
a molecule-like state in \cite{wzg15}. The above informations indicate that
the $Z_{c}(4200)$ is an ideal candidate for investigating the nature of
exotic charmoniumlike states.

As of now, the charmoniumlike $XYZ$ states are only observed in four ways
\cite{lx14}, i.e., the $e^{+}e^{-}$ annihilation ($e^{+}e^{-}\rightarrow XYZ$
or $e^{+}e^{-}\rightarrow J/\psi +XYZ$), $\gamma \gamma $ fusion process ($%
\gamma \gamma \rightarrow XYZ$), $B$ meson decay ($B\rightarrow K+XYZ$) and
hidden-charm dipion decays of higher charmonia or charmoniumlike states.
Therefore, searching for the charmoniumlike states through other production
process is an important topic, which will be useful in confirming and
understanding these exotic $XYZ$ states. For example, Ke \textit{et al. }%
suggested to search for the charged $Z_{c}^{\pm }(4430)$ by the
nucleon-antinucleon scattering \cite{ke08}, while the production of neutral $%
Z_{c}^{0}(4430)$ and $Z_{c}^{0}(4200)$ states in $\bar{p}p$ reaction were
investigated in Refs. \cite{wang15,ahep15}. Moreover, in Refs. \cite%
{lxh08,he09,lin13,lin14}, the meson photoproduction process were proposed to
be an effective way to search for the charmoniumlike states. Soon after,
according to the theoretical predictions obtained in Ref. \cite{lin13}, an
experiment of searching for the $Z_{c}^{\pm }(3900)$ through $\gamma
N\rightarrow Z_{c}^{\pm }(3900)N\rightarrow J/\psi \pi ^{\pm }N$ was carried
out by the COMPASS Collaboration \cite{compass}. Unfortunately, no signal of
exclusive photoproduction of the $Z_{c}^{\pm }(3900)$ state and its decay
into $J/\psi \pi ^{\pm }$ was found. Thus it is important to discuss whether
there are other charmoniumlikes that have a discovery potential through $%
\gamma N\rightarrow J/\psi \pi ^{\pm }N$ channel. Besides, a more accuracy
theoretical prediction is necessary.

Usually, for the meson photoproduction process, the mesonic Reggeized
treatment will play important role at high photo energies. The exchange of
dominant meson Regge trajectories already used to successfully describe the
meson photoproduction in Refs. \cite{mg97,gg11,he14}. Since a high photon
beam energy is required for the production of charmoniumlike states through
meson photoproduction process, the Reggeized treatment will be necessary to
ensure the result accuracy. In this work, within the frame of an effective
Lagrangian approach and the Regge trajectories model, we systematically
study the production of charged $Z_{c}(4200)$ by meson photoproduction
process in order to provide a reliable theoretical results and shed light on
our understanding of the properties and production mechanism of charged $%
Z_{c}(4200)$ state.

This paper is organized as follows. After an introduction, we present the
investigate method and formalism. The numerical result and discussion are
given in Sec. III. In Sec. IV, we discuss the upper limit of decay width of $%
Z_{c}(4200)\rightarrow J/\psi \pi $. Finally, this paper ends with a brief
conclusion.

\section{Formalism and ingredients}

Since the $Z_{c}(4200)$ have a strong coupling with $J/\psi \pi $ \cite%
{belle14,zl14,wei15}, the photoproduction process $\gamma p\rightarrow
Z_{c}^{+}(4200)n\rightarrow J/\psi \pi ^{+}n$ may be an ideal reaction
channel of searching and studying production of the charged $Z_{c}^{+}(4200)$%
. Moreover, considering the signal of $Z_{c}^{+}(4200)$ are mainly from the
contributions of $\pi $ exchange, while the contributions from $\rho $ and $%
a_{0}$ exchange can be negligible\footnote{%
In Refs. \cite{ab06,va06,sc02}, the results indicate that the pion exchange
plays a major role in the $\gamma p\rightarrow Xn$ process by analyzing the
HERA data. Besides, In Refs. \cite{bz96,hh95}, it is found that the
contributions of $\rho $ and $a_{0}$ exchange in the $\gamma ^{\ast
}p\rightarrow Xn$ reaction are very small. Thus, in the present work we only
consider the contribution from the one pion exchange. Here, the $\gamma
^{\ast }$ stand for the virtual photon.}, the process as depicted in Fig. 1
are regard as the source of signal of $Z_{c}^{+}(4200)$. Besides, the
reaction $\gamma p\rightarrow J/\psi \pi ^{+}n$ via Pomeron exchange (as
shown in Fig.2) are also calculated, which is considered to be the
background for the $Z_{c}^{+}(4200)$ photoproduction. To investigate $%
Z_{c}^{+}(4200)$ production, an effective Lagrangian approach and the Regge
trajectories model in terms of hadrons will be used in the follows.

\subsection{Feynman diagrams and effective interaction Lagrangian densities}

\begin{figure}[bp]
\centering
\includegraphics[scale=0.6]{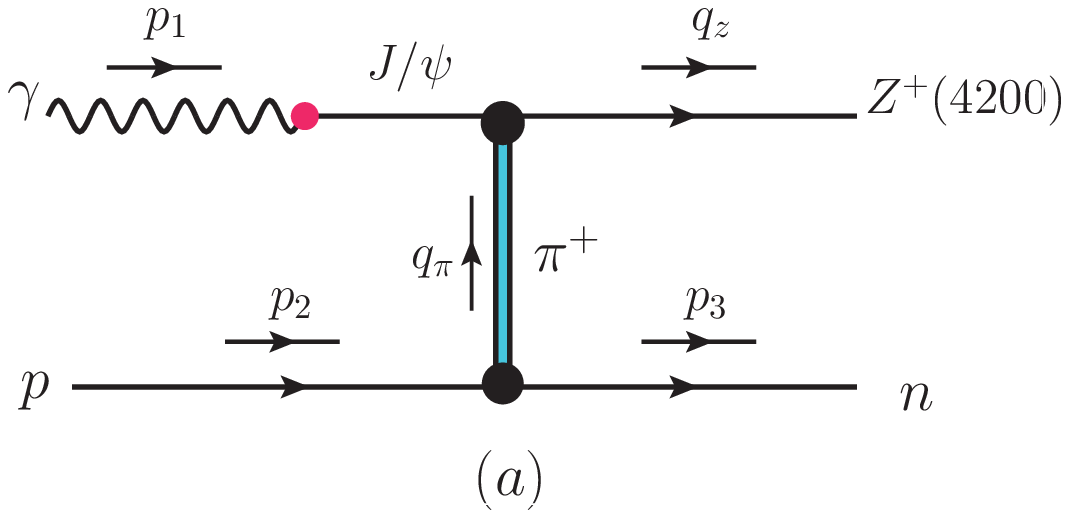} \hspace{80pt} %
\includegraphics[scale=0.6]{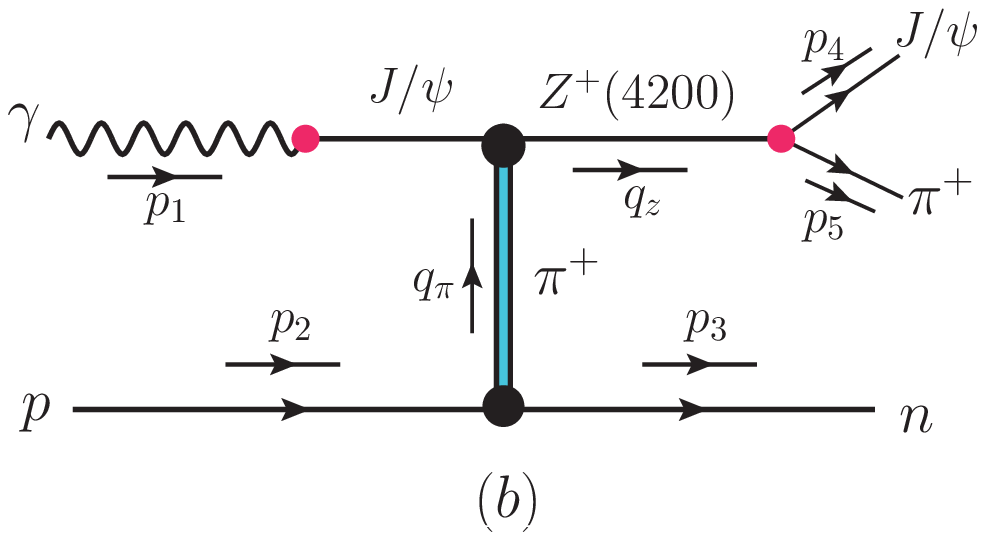}
\caption{(Color online) The Feynman diagram for $\protect\gamma p\rightarrow
Z_{c}^{+}(4200)n$ reaction (a) and $\protect\gamma p\rightarrow J/\protect%
\psi \protect\pi ^{+}n$ reaction (b) through $\protect\pi $ exchange. }
\end{figure}

Fig. 1 show the basic tree level Feynman diagram for the production of $%
Z_{c}^{+}(4200)$ in $\gamma p\rightarrow Z_{c}^{+}(4200)n\rightarrow J/\psi
\pi ^{+}n$ reaction via pion exchange. To gauge the contributions of these
diagrams, we need to know the effective Lagrangian densities for each
interaction vertex.

For the interaction vertex of $\pi NN$, we use the effective pseudoscalar
coupling \footnote{%
It should be noted that some works \cite{mb13,ys15} have pointed out that
the simple pseudoscalar coupling between nucleons and pions is incomplete
and inconsistent with chiral symmetry. Thus the pseudovector coupling is
suggested in Refs. \cite{mb13,ys15}. However, since the new pseudovector
formalism may not yet be ready for phenomenological use \cite{fc}, the
pseudoscalar coupling is adopted in the present work.} \cite{kt94,kt97,kt98},%
\begin{equation}
\mathcal{L}_{\pi NN}=-ig_{\pi NN}\bar{N}\gamma _{5}\vec{\tau}\cdot \vec{\pi}N
\end{equation}%
where $N$ and $\pi $ stand for the fields of nucleon and pion meson, while $%
\vec{\tau}$ is the Pauli matrix. The coupling constant of the $\pi NN$
interaction was given in many theoretical works, and we take $g_{\pi
NN}^{2}/4\pi =14.4$ \cite{lzw}.

\begin{figure}[t]
\begin{center}
\includegraphics[scale=0.6]{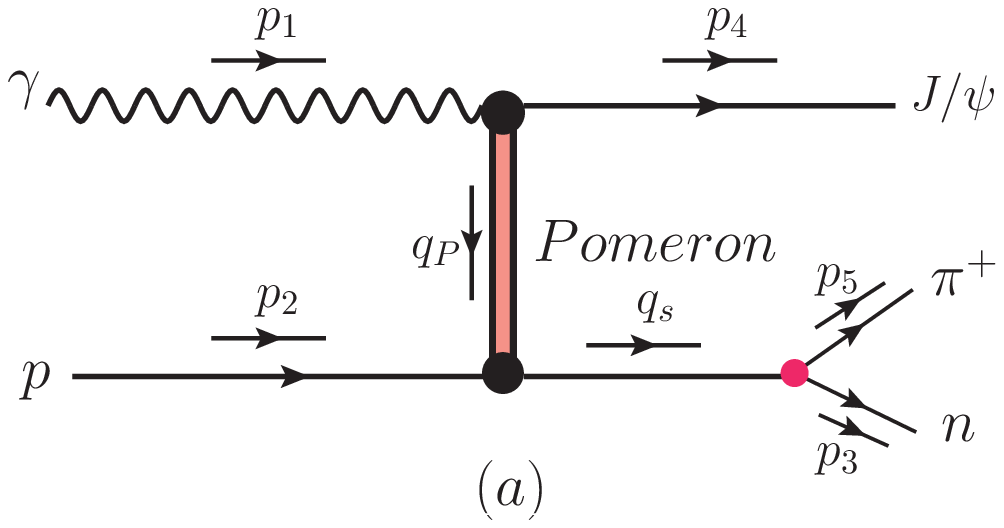} \hspace{80pt} %
\includegraphics[scale=0.6]{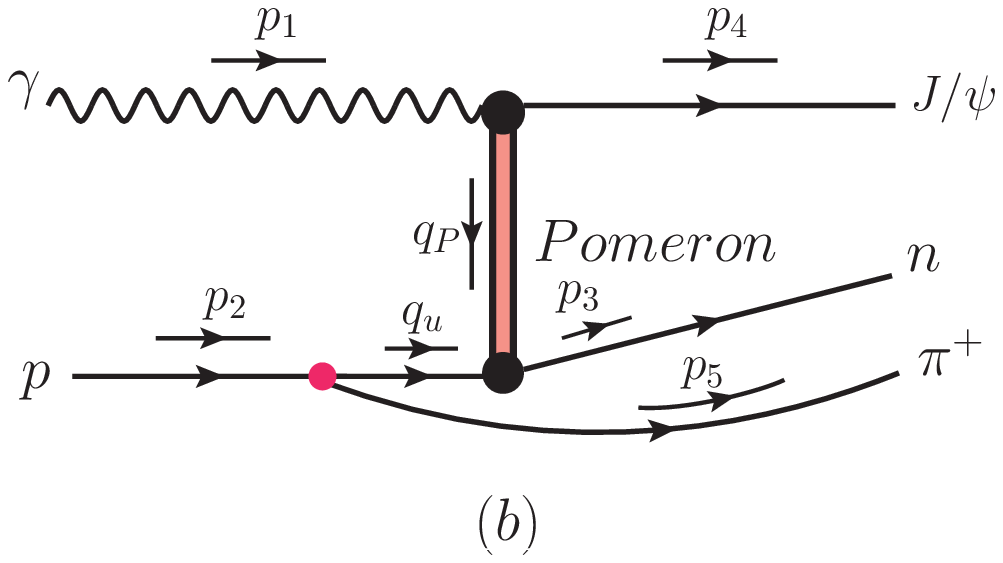}
\end{center}
\caption{(Color online) The Feynman diagram of $\protect\gamma p\rightarrow
J/\protect\psi \protect\pi ^{+}n$ process through the Pomeron exchange. }
\label{ozi}
\end{figure}

As mentioned above, the spin-parity of $Z_{c}^{+}(4200)$ has been determined
by Belle Collaboration to be $J^{P}=1^{+}$ \cite{belle14}. Thus the relevant
effective Lagrangian for the vertex\footnote{%
For the sake of simplicity, we use $Z$ and $\psi $ denote $Z_{c}(4200)$ and $%
J/\psi $, respectively.} of $Z\psi \pi $ read as \cite{lxh08},%
\begin{equation}
\mathcal{L}_{Z\psi \pi }=\frac{g_{Z\psi \pi }}{M_{Z}}(\partial ^{\mu }\psi
^{\nu }\partial _{\mu }\pi Z_{\nu }-\partial ^{\mu }\psi ^{\nu }\partial
_{\nu }\pi Z_{\mu }),
\end{equation}%
where $Z$, and $\psi $ denote the fields of $Z(4200)$ and $J/\psi $ meson,
respectively. With the effective Lagrangians above, the coupling constant $%
g_{Z\psi \pi }$ can be determined by the partial decay widths $\Gamma
_{Z_{c}(4200)\rightarrow J/\psi \pi }$,

\begin{eqnarray}
\Gamma _{Z(4200)\rightarrow J/\psi \pi } &=&\left( \frac{g_{Z\psi \pi }}{%
M_{Z}}\right) ^{2}\frac{|\vec{p}_{\pi }^{~\mathrm{c.m.}}|}{24\pi M_{Z}^{2}}
\nonumber \\
&&\times \left[ \frac{(M_{Z}^{2}-m_{\psi }^{2}-m_{\pi }^{2})^{2}}{2}+m_{\psi
}^{2}E_{\pi }^{2}\right] ,
\end{eqnarray}%
with%
\begin{eqnarray}
|\vec{p}_{\pi }^{~\mathrm{c.m.}}| &=&\frac{\lambda ^{1/2}(M_{Z}^{2},m_{\psi
}^{2},m_{\pi }^{2})}{2M_{Z}}, \\
E_{\pi } &=&\sqrt{|\vec{p}_{\pi }^{~\mathrm{c.m.}}|^{2}+m_{\pi }^{2}},
\end{eqnarray}%
where $\lambda $ is the K$\ddot{a}$llen function with $\lambda
(x,y,z)=(x-y-z)^{2}-4yz$.

As of now, no relevant experiment datas about $\Gamma _{Z(4200)\rightarrow
J/\psi \pi }$ can be available \cite{pdg14}. However, in Ref. \cite{wei15},
the authors obtained the partial decay width $\Gamma
_{Z_{c}(4200)\rightarrow J/\psi \pi }=87.3\pm 47.1$ MeV with the QCD sum
rule approach, which allow us to estimate the lower (upper) limit of the
decay width of $Z(4200)\rightarrow J/\psi \pi $, $\Gamma
_{Z(4200)\rightarrow J/\psi \pi }=40.2(134.4)$ MeV. With $M_{Z}=4196$ MeV/$%
c^{2}$ and $\Gamma _{Z}=370$ MeV \cite{pdg14}, we get $g_{Z\psi \pi
}/M_{z}=1.174,1.731,2.147$ MeV, which correspond to three typical partial
decay width $\Gamma _{Z(4200)\rightarrow J/\psi \pi }=40.2,87.3,134.4$ MeV,
respectively.

For the interaction vertex of $Z\gamma \pi $, we need to derive it by the
vector meson dominance (VMD) mechanism \cite{tb65,tb69,tb79}. In the VMD
mechanism for photoproduction, a real photon can fluctuate into a virtual
vector meson, which subsequently scatters off the target proton. Thus within
the frame of VMD mechanism, we get the Lagrangian of depicting the coupling
of the intermediate vector meson $J/\psi $ with a photon as follows,%
\begin{equation}
\mathcal{L}_{J/\psi \gamma }=-\frac{em_{\psi }^{2}}{f_{\psi }}V_{\mu }A^{\mu
},
\end{equation}%
where $m_{\psi }^{2}$ and $f_{\psi }$ are the mass and the decay constant of
$J/\psi $ meson, respectively. With the above equation, one gets the
expression for the $J/\psi \rightarrow e^{+}e^{-}$ decay,%
\begin{equation}
\Gamma _{J/\psi \rightarrow e^{+}e^{-}}=\left( \frac{e}{f_{\psi }}\right)
^{2}\frac{8\alpha \left\vert \vec{p}_{e}^{~\mathrm{c.m.}}\right\vert ^{3}}{%
3m_{\psi }^{2}},
\end{equation}%
where $\vec{p}_{e}^{~\mathrm{c.m.}}$ indicate the three-momentum of an
electron in the rest frame of the $J/\psi $ meson, while $\alpha
=e^{2}/4\hbar c=1/137$ is the electromagnetic fine structure constant. Thus,
in the light of the partial decay width of $J/\psi \rightarrow e^{+}e^{-}$
\cite{pdg14}%
\begin{equation}
\Gamma _{J/\psi \rightarrow e^{+}e^{-}}\simeq 5.547\text{ keV,}
\end{equation}%
we get the constant $e/f_{\psi }\simeq 0.027$.

In Fig. 2, we present the Feynman diagram for the $\gamma p\rightarrow
J/\psi \pi ^{+}n$ process through Pomeron exchange, which is considered as
the main background contributions to $\gamma p\rightarrow
Z_{c}^{+}(4200)n\rightarrow J/\psi \pi ^{+}n$ process. To depict the Pomeron
exchange process, the relevant formulas which were used in Refs. \cite%
{lxh08,ad87,ma96} are adopted in this work. The Pomeron-nucleon coupling is
described as follow,%
\begin{equation}
F_{\mu }(t)=\frac{3\beta _{0}(4m_{N}^{2}-2.8t)}{(4m_{N}^{2}-t)(1-t/0.7)^{2}}%
\gamma _{\mu }=F(t)\gamma _{\mu }\text{,}
\end{equation}%
where $t=q_{P}^{2}$ is the exchanged Pomeron momentum squared. $\beta
_{0}^{2}=4$ GeV$^{2}$ stands for the coupling constant between a single
Pomeron and a light constituent quark.

For the vertex of $\gamma \psi \mathcal{P}$, with an on-shell approximation
for keeping the gauge invariance, we have%
\begin{equation}
V_{\gamma \psi \mathcal{P}}=\frac{2\beta _{c}\times 4\mu _{0}^{2}}{(m_{\psi
}^{2}-t)(2\mu _{0}^{2}+m_{\psi }^{2}-t)}T_{\mu \rho \nu }\epsilon _{\psi
}^{\nu }\epsilon _{\gamma }^{\mu }\mathcal{P}^{\rho },
\end{equation}%
with%
\begin{eqnarray}
T^{\mu \rho \nu } &=&(p_{1}+p_{4})^{\rho }g^{\mu \nu }-2p_{1}^{\nu }g^{\rho
\mu }  \nonumber \\
&&+2\Big\{p_{1}^{\mu }g^{\rho \nu }+\frac{p_{4}^{\nu }}{p_{4}^{2}}%
(p_{1}\cdot p_{4}g^{\rho \mu }-p_{1}^{\rho }p_{4}^{\mu }-p_{1}^{\mu
}p_{4}^{\rho })  \nonumber \\
&&-\frac{p_{1}^{2}p_{4}^{\mu }}{p_{4}^{2}p_{1}\cdot p_{4}}(p_{4}^{2}g^{\rho
\nu }-p_{4}^{\rho }p_{4}^{\nu })\Big\}+(p_{1}-p_{4})^{\rho }g^{\mu \nu },
\end{eqnarray}%
where $\beta _{c}^{2}=0.8$ GeV$^{2}$ is the effective coupling constant
between a Pomeron and a charm quark within $J/\psi $ meson, while $\mu
_{0}=1.2$ GeV$^{2}$ denotes a cutoff parameter in the form factor of Pomeron.

\subsection{Cross sections for the $\protect\gamma p\rightarrow
Z_{c}^{+}(4200)n$ reaction}

After the above preparations, the invariant scattering amplitude $\mathcal{A}
$ for the $\gamma (p_{1})p(p_{2})\rightarrow Z_{c}^{+}(4200)(q_{z})n(p_{3})$
reaction by exchanging a $\pi $ meson read as,%
\begin{eqnarray}
\mathcal{A} &=&(\sqrt{2}g_{\pi NN}\frac{g_{Z\psi \pi }}{M_{Z}}\frac{e}{%
f_{\psi }})\bar{u}(p_{3})\gamma _{5}u(p_{2})\epsilon _{Z}^{\ast \mu }
\nonumber \\
&&\times \epsilon _{\gamma }^{\nu }\left[ p_{1}\cdot (q_{z}-p_{1})g_{\mu \nu
}-p_{1\mu }(q_{z}-p_{1})_{\nu }\right]  \nonumber \\
&&\times \frac{1}{q_{\pi }^{2}-m_{\pi }^{2}}F_{\pi NN}(q_{\pi }^{2})F_{Z\psi
\pi }(q_{\pi }^{2})
\end{eqnarray}%
where $F_{\pi NN}(q_{\pi }^{2})$ and $F_{Z\psi \pi }(q_{\pi }^{2})$ are the
form factors for the vertices of $\pi NN$ and $Z\psi \pi $, respectively. We
have the following definitions for both form factors,%
\begin{equation}
F_{\pi NN}(q_{\pi }^{2})=\frac{\Lambda _{\pi }^{2}-m_{\pi }^{2}}{\Lambda
_{\pi }^{2}-q_{\pi }^{2}},
\end{equation}%
and%
\begin{equation}
F_{Z\psi \pi }(q_{\pi }^{2})=\frac{m_{\psi }^{2}-m_{\pi }^{2}}{m_{\psi
}^{2}-q_{\pi }^{2}},
\end{equation}%
where $\Lambda _{\pi }$ is the cutoff parameter for the $\pi NN$ vertex. In
the next calculations, we take the typical value of $\Lambda _{\pi }=0.7$
GeV as used in Refs. \cite{lxh08,lin13,gg11,vp14}.
\begin{figure}[b]
\centering
\includegraphics[scale=0.4]{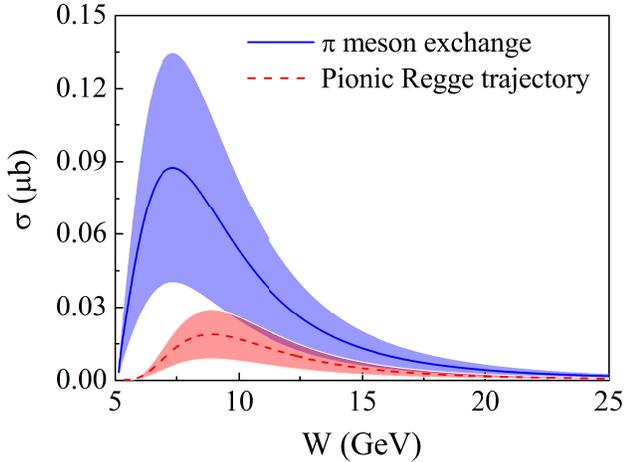}
\caption{(Color online) The total cross section for $\protect\gamma %
p\rightarrow Z_{c}^{+}(4200)n$ process through $\protect\pi $ meson or
pionic Regge trajectory exchange. Here, the numerical results (the blue
solid line and the red dashed line) correspond to the partial decay width $%
\Gamma _{Z_{c}(4200)\rightarrow J/\protect\psi \protect\pi }=87.3$ MeV,
while the bands stand for the uncertainties with the variation of $\Gamma
_{Z_{c}(4200)\rightarrow J/\protect\psi \protect\pi }$ from 40.2 to 131.4
MeV.}
\end{figure}

\begin{figure}[tbph]
\centering
\includegraphics[scale=0.4]{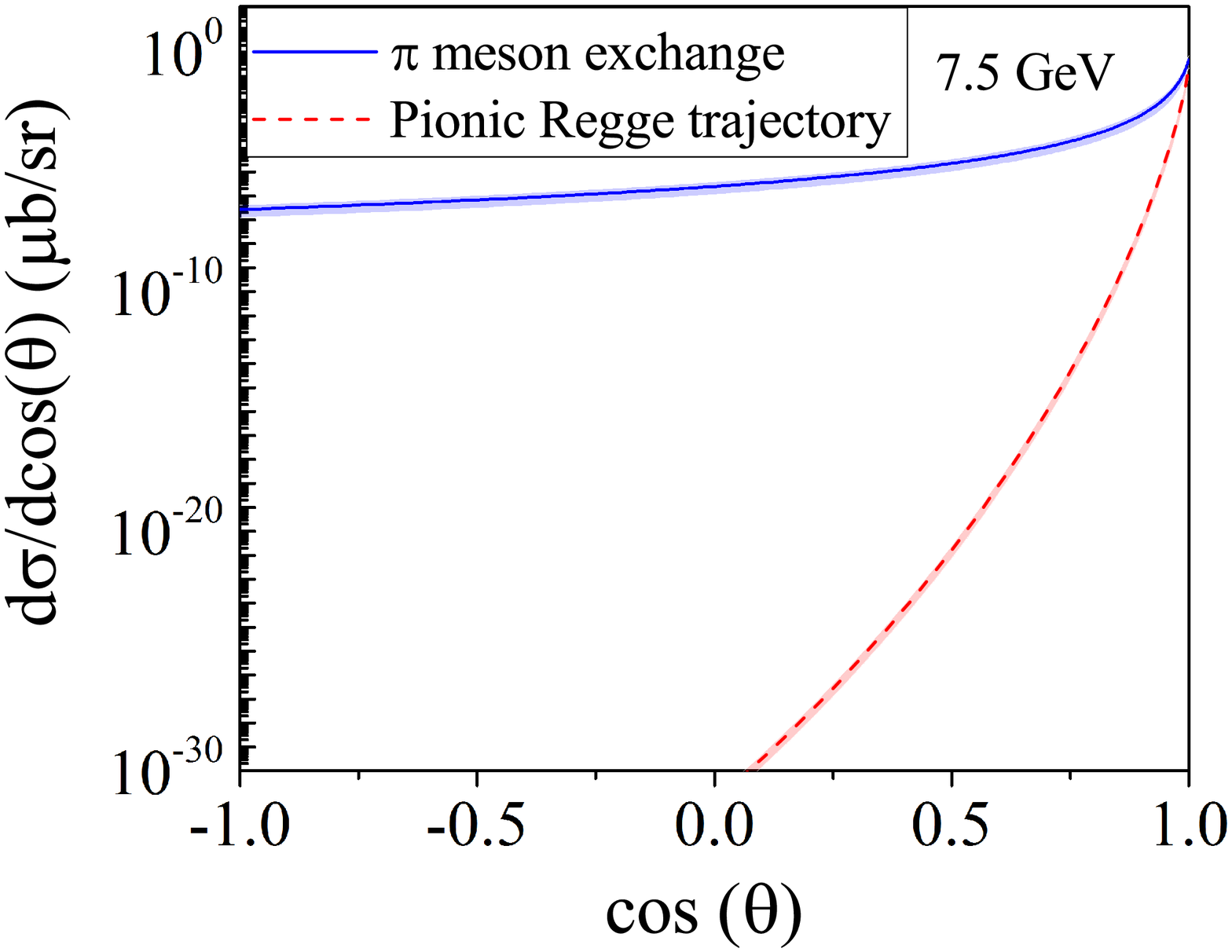} \hspace{80pt} %
\includegraphics[scale=0.4]{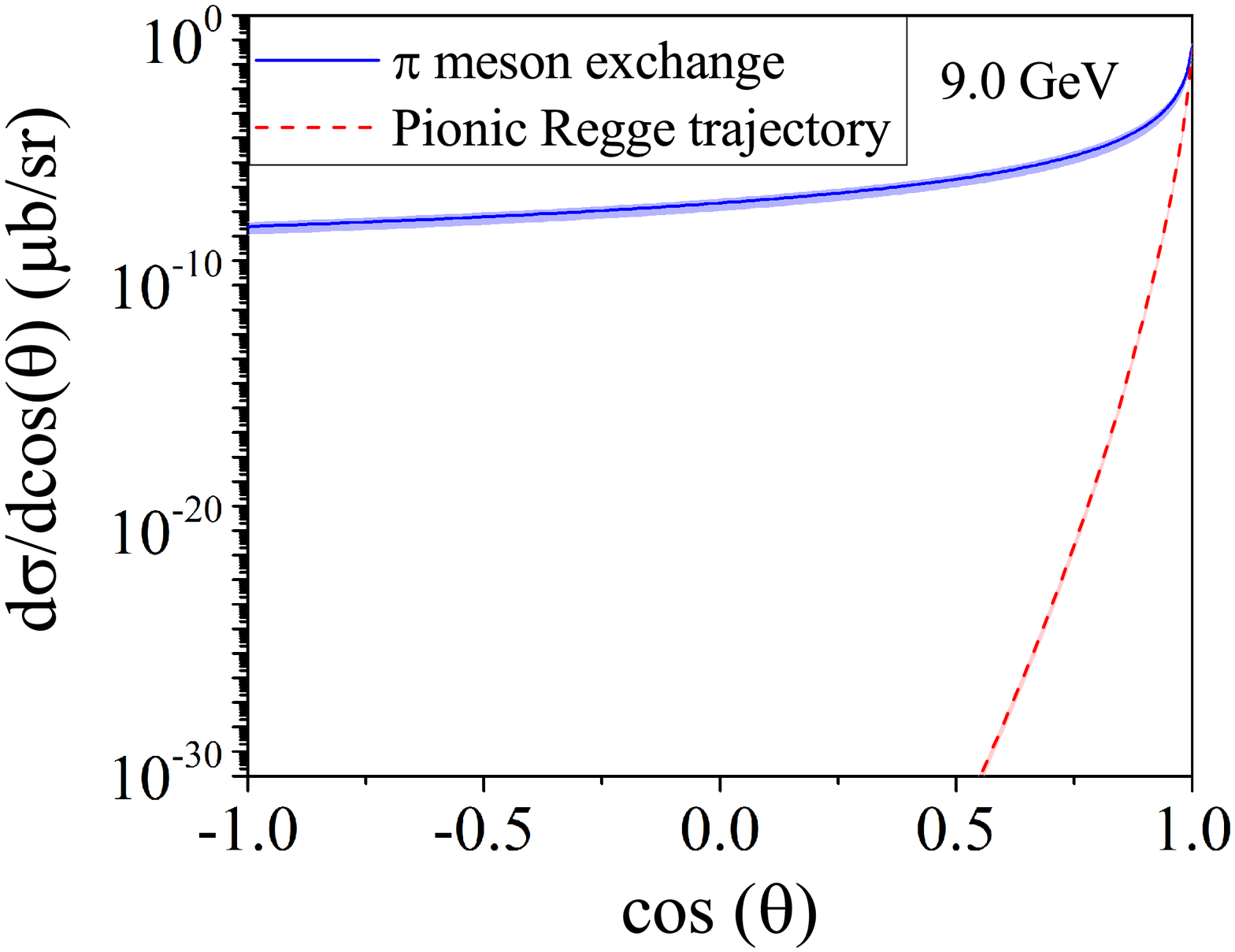}
\caption{(Color online) The differential cross section for $\protect\gamma %
p\rightarrow Z_{c}^{+}(4200)n$ process through $\protect\pi $ meson or
pionic Regge trajectory exchange. The notation of the lines and bands as in
Fig. 3.}
\end{figure}
As mentioned above, a higher photon beam energy is required for the
production of charmoniumlike states through meson photoproduction process.
Thus, to better describe the photoproduction of $Z_{c}^{+}(4200)$ at high
photon energies we introduce a pion Reggeized treatment by replacing the
Feynman propagator $\frac{1}{q_{\pi }^{2}-m_{\pi }^{2}}$ with the Regge
propagator as follows \cite{mg97,gg11,he14,rj71},%
\begin{equation}
\frac{1}{q_{\pi }^{2}-m_{\pi }^{2}}\rightarrow \mathcal{R}_{\pi }=(\frac{s}{%
s_{\text{scale}}})^{\alpha _{\pi }(t)}\frac{\pi \alpha _{\pi }^{\prime }}{%
\Gamma \lbrack 1+\alpha _{\pi }(t)]}\frac{e^{-i\pi \alpha _{\pi }(t)}}{\sin
[\pi \alpha _{\pi }(t)]},
\end{equation}%
where $\alpha _{\pi }^{\prime }$ is the slope of the trajectory and the
scale factor $s_{\text{scale}}$ is fixed at 1 GeV$^{2}$, while $%
s=(p_{1}+p_{2})^{2}$ and $t=(p_{2}+p_{3})^{2}$ are the Mandelstam variables.
In addition, the pionic Regge trajectory $\alpha _{\pi }(t)$ read as \cite%
{gg11,he14,rj71}%
\begin{equation}
\alpha _{\pi }(t)=0.7(t-m_{\pi }^{2}).
\end{equation}%
The unpolarized differential cross section for the $Z_{c}^{+}(4200)$
photoproduction shown in Fig. 1(a) then reads%
\begin{equation}
\frac{d\sigma }{d\cos \theta }=\frac{1}{32\pi s}\frac{\left\vert \vec{q}%
_{z}^{~\mathrm{c.m.}}\right\vert }{\left\vert \vec{p}_{1}^{~\mathrm{c.m.}%
}\right\vert }\left( \frac{1}{4}\sum\limits_{spins}\left\vert \mathcal{A}%
\right\vert ^{2}\right) ,
\end{equation}%
where $\vec{p}_{1}^{~\mathrm{c.m.}}$ and $\vec{q}_{z}^{~\mathrm{c.m.}}$ are
the three-momentum of initial photon and final $Z_{c}^{+}(4200)$ state,
while $\theta $ denotes the angle of the outgoing $Z_{c}^{+}(4200)$ state
relative to the photon beam direction in the c.m. frame. The total cross
section can be easily obtained by integrating the above equation.

In Fig. 3, the total cross section $\sigma (\gamma p\rightarrow Z_{c}^{+}n)$
through $\pi $ meson or pionic Regge trajectory exchange are presented with $%
\Lambda _{\pi }=0.7$ GeV. Since the total cross section is proportional to
the partial decay width $\Gamma _{Z(4200)\rightarrow J/\psi \pi }$, thus we
note that the cross section changes by a factor of 3 to 4 when the partial
width $\Gamma _{Z(4200)\rightarrow J/\psi \pi }$ is varied from $40.2$ to $%
134.4$ MeV. Besides, it is found that the total cross section through the
Reggeized treatment is about five times smaller than that of result through
a $\pi $ exchange, which indicate that the Reggeized treatment can lead to a
lower cross section of the $Z_{c}^{+}(4200)$ photoproduction at high photon
energies. Moreover, we note that the peak position of total cross section
was moved to the higher energy point when the Reggeized treatment is used in
the calculations.

Fig. 4 are the differential cross section for the $\gamma p\rightarrow
Z_{c}^{+}n$ process by exchanging the $\pi $ meson or pionic Regge
trajectory at different energies, respectively. From Fig. 4 one can see
that, relative to the results related to the $\pi $ exchange, the
differential cross section by exchanging the pionic Regge trajectory are
very sensitive to the $\theta $ angle and gives a considerable contribution
at forward angles.

\subsection{Cross sections for the $\protect\gamma p\rightarrow J/\protect%
\psi \protect\pi ^{+}n$ reaction}

With the Feynman rules and above Lagrangian densities, we obtain the
invariant scattering amplitude $\mathcal{M}_{Z}^{signal}$ for the $\gamma
p\rightarrow J/\psi \pi ^{+}n$ process through $\pi $ exchange (as depicted
in Fig. 1(b)) as follows,%
\begin{eqnarray}
\mathcal{M}_{Z}^{signal} &\mathcal{=}&(\sqrt{2}g_{\pi NN}\frac{g_{Z\psi \pi }%
}{M_{Z}}\frac{e}{f_{\psi }})\bar{u}(p_{3})\gamma _{5}u(p_{2})  \nonumber \\
&&\times (p_{1}\cdot q_{\pi }g^{\mu \alpha }-p_{1}^{\alpha }q_{\pi }^{\mu
})(p_{4}\cdot p_{5}g^{\rho \nu }-p_{4}^{\rho }p_{5}^{\nu })  \nonumber \\
&&\times \frac{1}{q_{\pi }^{2}-m_{\pi }^{2}}G_{Z}^{\rho \alpha
}(q_{z})\epsilon _{\gamma \mu }\epsilon _{\psi \nu }^{\ast }\left( \frac{%
\Lambda _{\pi }^{2}-m_{\pi }^{2}}{\Lambda _{\pi }^{2}-q_{\pi }^{2}}\right)
\nonumber \\
&&\times \left( \frac{m_{\psi }^{2}-m_{\pi }^{2}}{m_{\psi }^{2}-q_{\pi }^{2}}%
\right) \left( \frac{m_{\psi }^{2}-M_{Z}^{2}}{m_{\psi }^{2}-q_{z}^{2}}%
\right) ,
\end{eqnarray}%
where $G_{Z}^{\mu \alpha }$ are the propagators of the $Z(4200),$ taking the
Breit-Wigner form \cite{wh02},%
\begin{equation}
G_{Z}^{\rho \alpha }(q)=\frac{-g_{\rho \alpha }+q_{z\rho }q_{z\alpha
}/M_{Z}^{2}}{q_{z}^{2}-M_{Z}^{2}+iM_{Z}\Gamma _{Z}}.
\end{equation}%
Just as above practice, by replacing the Feynman propagator $\frac{1}{q_{\pi
}^{2}-m_{\pi }^{2}}$ with the Regge propagator $\mathcal{R}_{\pi }$, we can
get the scattering amplitude for the $\gamma p\rightarrow J/\psi \pi ^{+}n$
process through the pionic Regge trajectory exchange.

Since the Pomeron can mediate the long-range interaction between a confined
quark and a nucleon, thus $\gamma p\rightarrow J/\psi \pi ^{+}n$ via the
Pomeron exchange (as described in Fig. 2) are the mainly background
contribution to the $\gamma p\rightarrow Z_{c}^{+}(4200)n\rightarrow J/\psi
\pi ^{+}n$ reaction. The invariant scattering amplitudes $\mathcal{M}%
_{P}^{s} $ and $\mathcal{M}_{P}^{u}$ for Figs. 2(a) and 2(b) can be written,
respectively, as%
\begin{eqnarray}
\mathcal{M}_{P}^{s} &=&8\sqrt{2}\beta _{c}\mu _{0}^{2}g_{\pi
NN}F_{N}(q_{s}^{2})\frac{F(t)G_{P}(s,t)}{(m_{\psi }^{2}-t)(2\mu
_{0}^{2}+m_{\psi }^{2}-t)}  \nonumber \\
&&\times T^{\mu \rho \nu }\epsilon _{\psi \nu }^{\ast }\epsilon _{\gamma \mu
}\bar{u}(p_{3})\gamma _{5}\frac{\rlap{$\slash$}q_{s}+m_{N}}{%
q_{s}^{2}-m_{N}^{2}}\gamma _{\rho }u(p_{2}),
\end{eqnarray}%
\begin{eqnarray}
\mathcal{M}_{P}^{u} &=&8\sqrt{2}\beta _{c}\mu _{0}^{2}g_{\pi
NN}F_{N}(q_{u}^{2})\frac{F(t)G_{P}(s,t)}{(m_{\psi }^{2}-t)(2\mu
_{0}^{2}+m_{\psi }^{2}-t)}  \nonumber \\
&&\times T^{\mu \rho \nu }\epsilon _{\psi \nu }^{\ast }\epsilon _{\gamma \mu
}\bar{u}(p_{3})\gamma _{\rho }\frac{\rlap{$\slash$}q_{u}+m_{N}}{%
q_{u}^{2}-m_{N}^{2}}\gamma _{5}u(p_{2})
\end{eqnarray}%
with
\begin{equation}
G_{P}(s,t)=-i(\eta ^{\prime }s)^{\eta (t)-1}
\end{equation}%
where $\eta (t)=1+\epsilon +\eta ^{\prime }t$ is the Pomeron trajectory.
Here, the concrete values $\epsilon =0.08$ and $\eta ^{\prime }=0.25$ GeV$%
^{-2}$ are adopted.

Considering the size of the hadrons, the monopole form factor for the
off-shell intermediate nucleon is introduced as in the Bonn potential model
\cite{rm87}:%
\begin{equation}
F_{N}(q_{i}^{2})=\frac{\Lambda _{N}^{2}-m_{N}^{2}}{\Lambda _{N}^{2}-q_{i}^{2}%
},\text{ \ }i=s,u
\end{equation}%
where $\Lambda _{N}$ and $q_{i}(q_{s}=p_{3}+p_{5},q_{u}=p_{2}-p_{5})$ are
the cut-off parameter and four-momentum of the intermediate nucleon,
respectively. For the value of $\Lambda _{N}$, we will discuss it in the
next section. It is worth mentioning that the form factor is
phenomenological and has a great uncertainty. Thus the dipole form factor is
deserved to be discussed and compared with the monopole form.

Combining the signal terms and background amplitudes, we get the total
invariant amplitude%
\begin{equation}
\mathcal{M}=\mathcal{M}_{Z}^{signal}+\mathcal{M}_{P}^{s}+\mathcal{M}_{P}^{u}.
\end{equation}

Thus the total cross section of the $\gamma p\rightarrow J/\psi \pi ^{+}n$
reaction could be obtained by integrating the invariant amplitudes in the
three body phase space,%
\begin{eqnarray}
d\sigma (\gamma p &\rightarrow &J/\psi \pi ^{+}n)=\frac{m_{N}^{2}}{%
\left\vert p_{1}\cdot p_{2}\right\vert }\left( \frac{1}{4}%
\sum\limits_{spins}\left\vert \mathcal{M}\right\vert ^{2}\right)  \nonumber
\\
&&\times (2\pi )^{4}d\Phi _{3}(p_{1}+p_{2};p_{3},p_{4},p_{5}),
\end{eqnarray}%
where the three-body phase space is defined as \cite{pdg14}%
\begin{equation}
d\Phi _{3}(p_{1}+p_{2};p_{3},p_{4},p_{5})=\delta ^{4}\left(
p_{1}+p_{2}-\sum\limits_{i=3}^{5}p_{i}\right) \prod\limits_{i=3}^{5}\frac{%
d^{3}p_{i}}{(2\pi )^{3}2E_{i}}.
\end{equation}

\section{Numerical results and discussion}

With the FOWL code in the CERN program library, the total cross section
including both signal and background contributions can be calculated. In
these calculations, the cutoff parameters $\Lambda _{N}$ related to the
Pomeron term is a free parameter. Thus we first need to give a constraint on
the value of $\Lambda _{N}$. Fig. 5 (a) and Fig. 5 (b) present the variation
of cross section from the background contributions for $\gamma p\rightarrow
J/\psi \pi ^{+}n$ with monopole and dipole form factor, respectively. It is
obvious that the Pomeron exchange contributions with dipole form factor are
more sensitive to the values of the cutoff $\Lambda _{N}$ than that of
monopole form factor. Thus monopole form factor is adopted in the following
calculation.

At present, no experiment data is available for the $\gamma p\rightarrow
J/\psi \pi ^{+}n$ process. However, we notice that the similar reaction $%
\gamma p\rightarrow J/\psi p$ and $\bar{p}p\rightarrow J/\psi \pi ^{0}$ have
been measured by some experiment \cite{chu,al,E760,E835}, where the measured
cross sections of these two process are about 1 nb and 10 nb, respectively.
Here, we naively think that the cross section of $\gamma p\rightarrow J/\psi
\pi ^{+}n$ may be equal to or less than that of $\gamma p\rightarrow J/\psi
p $, while it greater than that of $\bar{p}p\rightarrow J/\psi \pi ^{0}$.
Thus we constrain the cutoff to be $\Lambda _{N}=0.96$ GeV as used in Ref.
\cite{lin13,lin14}, which will be used in our calculations.

\begin{figure}[tp]
\centering
\includegraphics[scale=0.4]{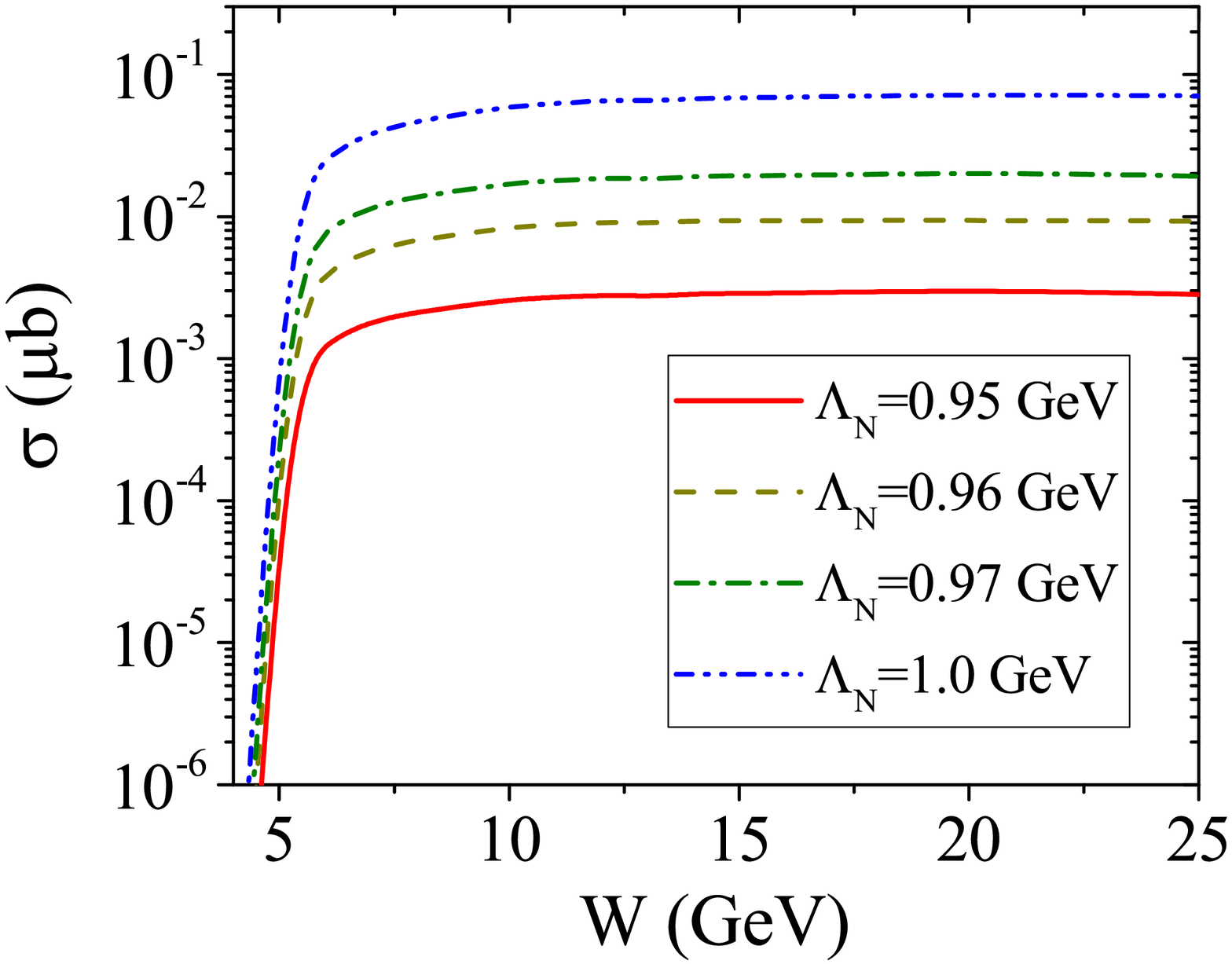} \includegraphics[scale=0.4]{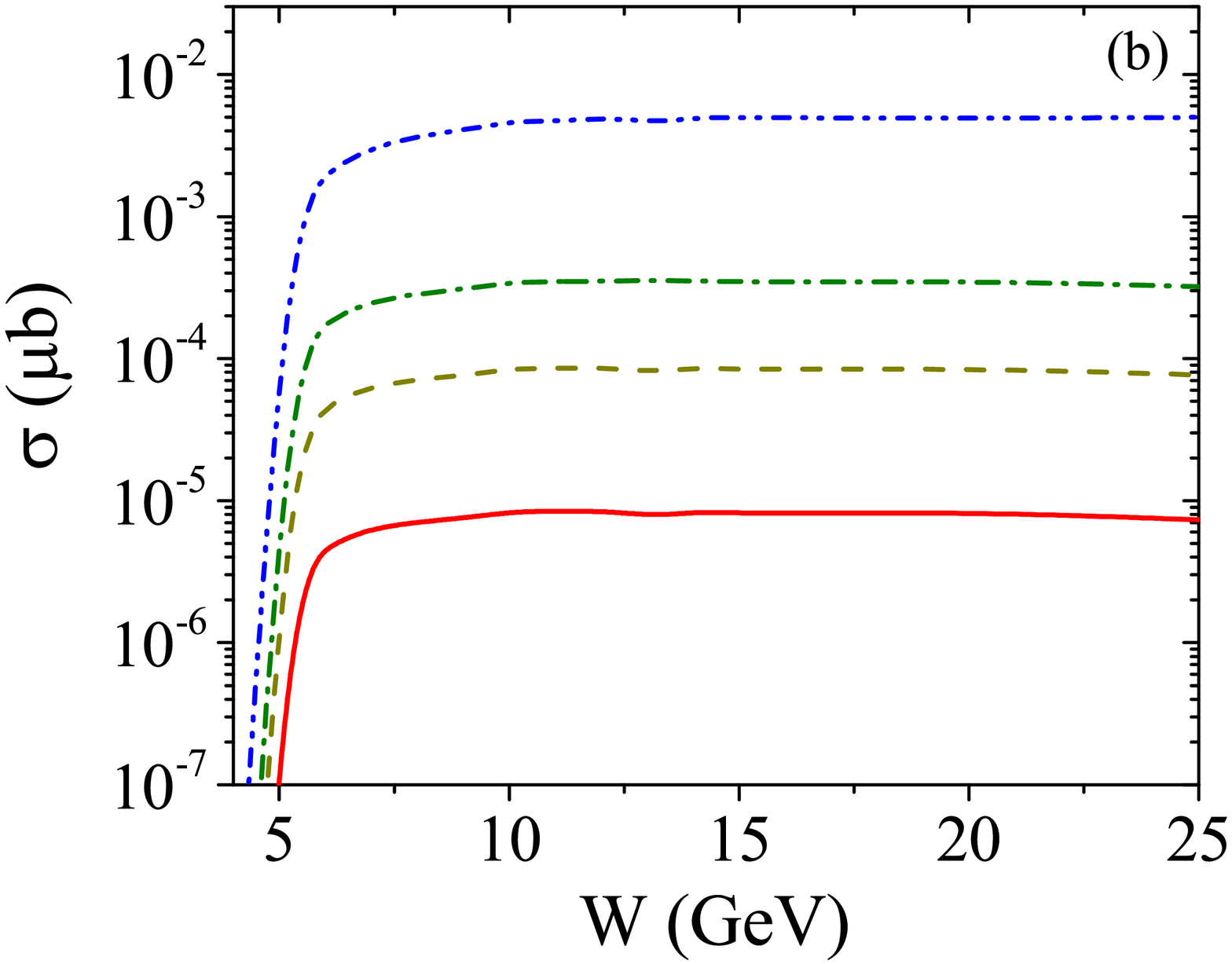}
\caption{(Color online) (a): The cross section of background from the
Pomeron exchange for $\protect\gamma p\rightarrow J/\protect\psi \protect\pi %
^{+}n$ process with monopole form factor at the different values of the
cutoff parameter $\Lambda _{N}$. (b) is same as the (a), but for the case of
dipole form factor.}
\end{figure}

\begin{figure}[bp]
\centering
\includegraphics[scale=0.4]{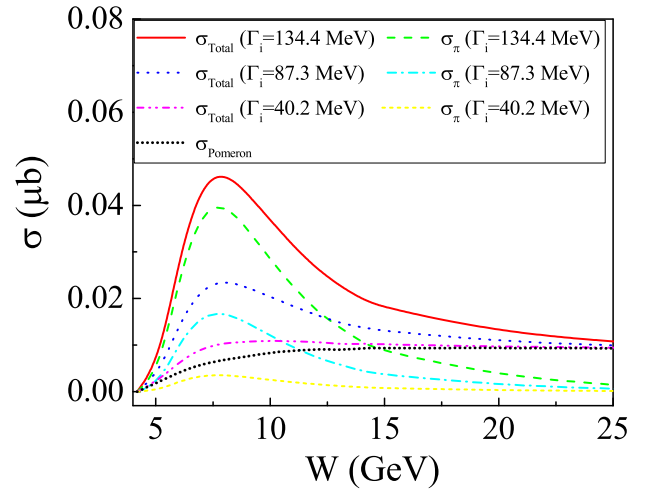}
\caption{(Color online) The energy dependence of the total cross sections
for the $\protect\gamma p\rightarrow J/\protect\psi \protect\pi ^{+}n$
reaction. Here, $\protect\sigma _{Pomeron}$ and $\protect\sigma _{\protect%
\pi }$ denote the results via the Pomeron and $\protect\pi $ exchange,
respectively, while $\protect\sigma _{Total}$ is the total cross section of $%
\protect\gamma p\rightarrow J/\protect\psi \protect\pi ^{+}n$. The variation
of $\protect\sigma _{\protect\pi }$ and $\protect\sigma _{Total}$ to $W$
with several typical partial width values $\Gamma _{Z(4200)\rightarrow J/%
\protect\psi \protect\pi }=40.2,87.3,134.4$ MeV are also presented.}
\end{figure}

\begin{figure}[tbph]
\centering
\includegraphics[scale=0.4]{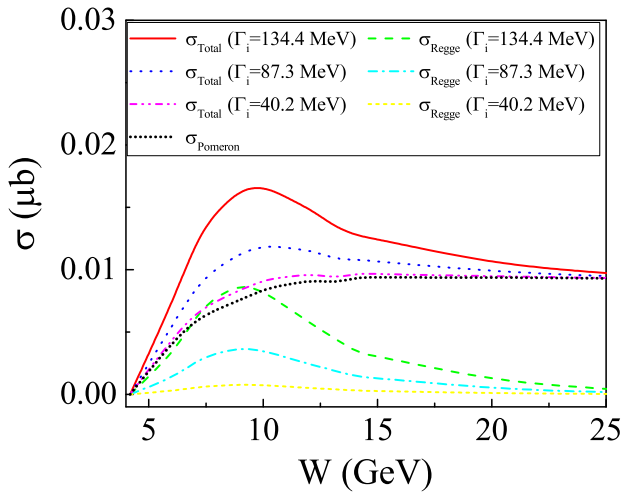}
\caption{(Color online) The energy dependence of the total cross sections
for the $\protect\gamma p\rightarrow J/\protect\psi \protect\pi ^{+}n$
reaction. Here, $\protect\sigma _{Pomeron}$ and $\protect\sigma _{Regge}$
denote the results via the Pomeron exchange and pionic Regge trajectory
exchange, respectively, while $\protect\sigma _{Total}$ is the total cross
section of $\protect\gamma p\rightarrow J/\protect\psi \protect\pi ^{+}n$.
The variation of $\protect\sigma _{Regge}$ and $\protect\sigma _{Total}$ to $%
W$ with several typical partial width values $\Gamma _{Z(4200)\rightarrow J/%
\protect\psi \protect\pi }=40.2,87.3,134.4$ MeV are also presented.}
\end{figure}

\begin{figure*}[t]
\begin{minipage}{1\textwidth}
\includegraphics[scale=0.29]{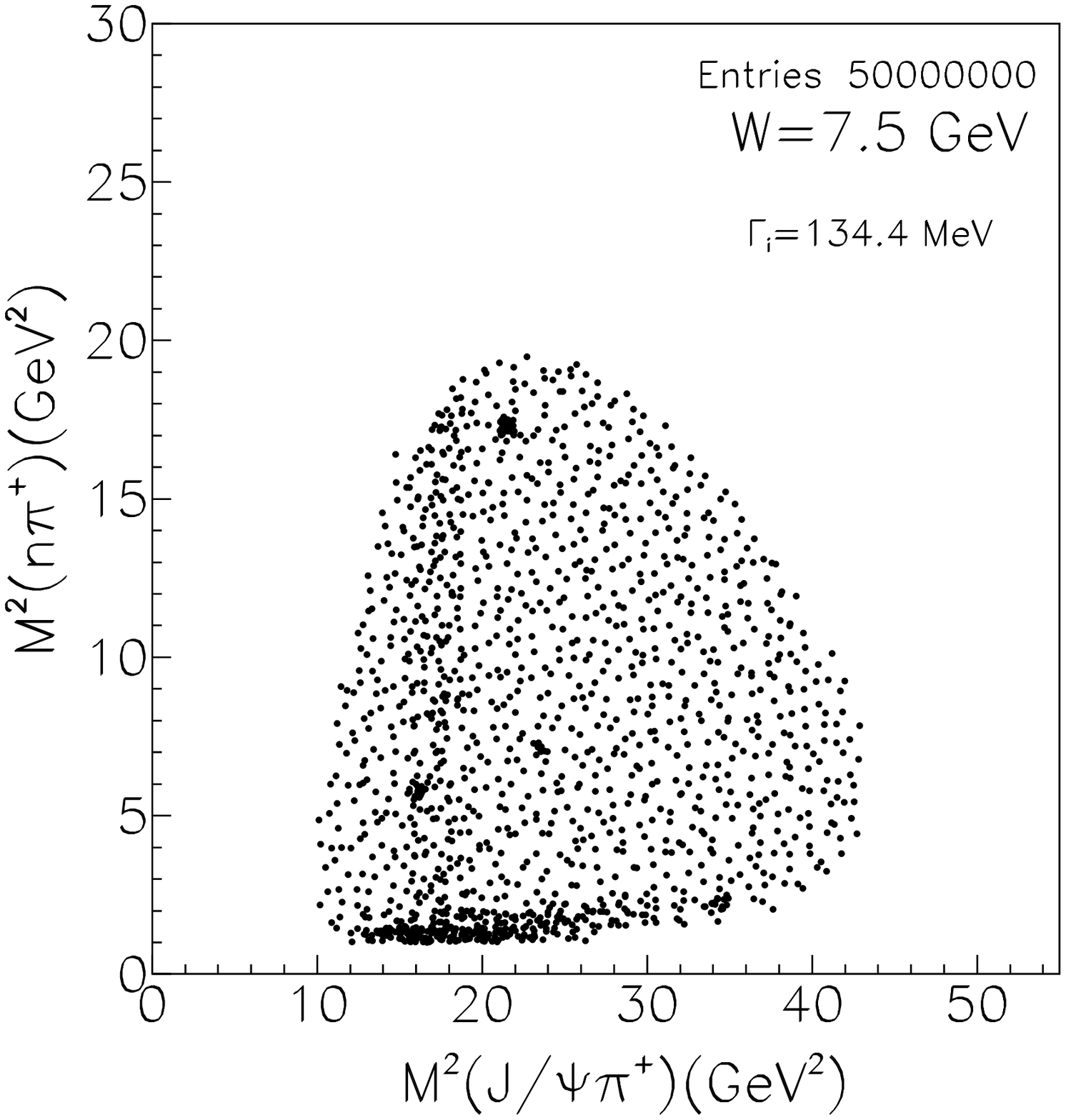}
\includegraphics[scale=0.29]{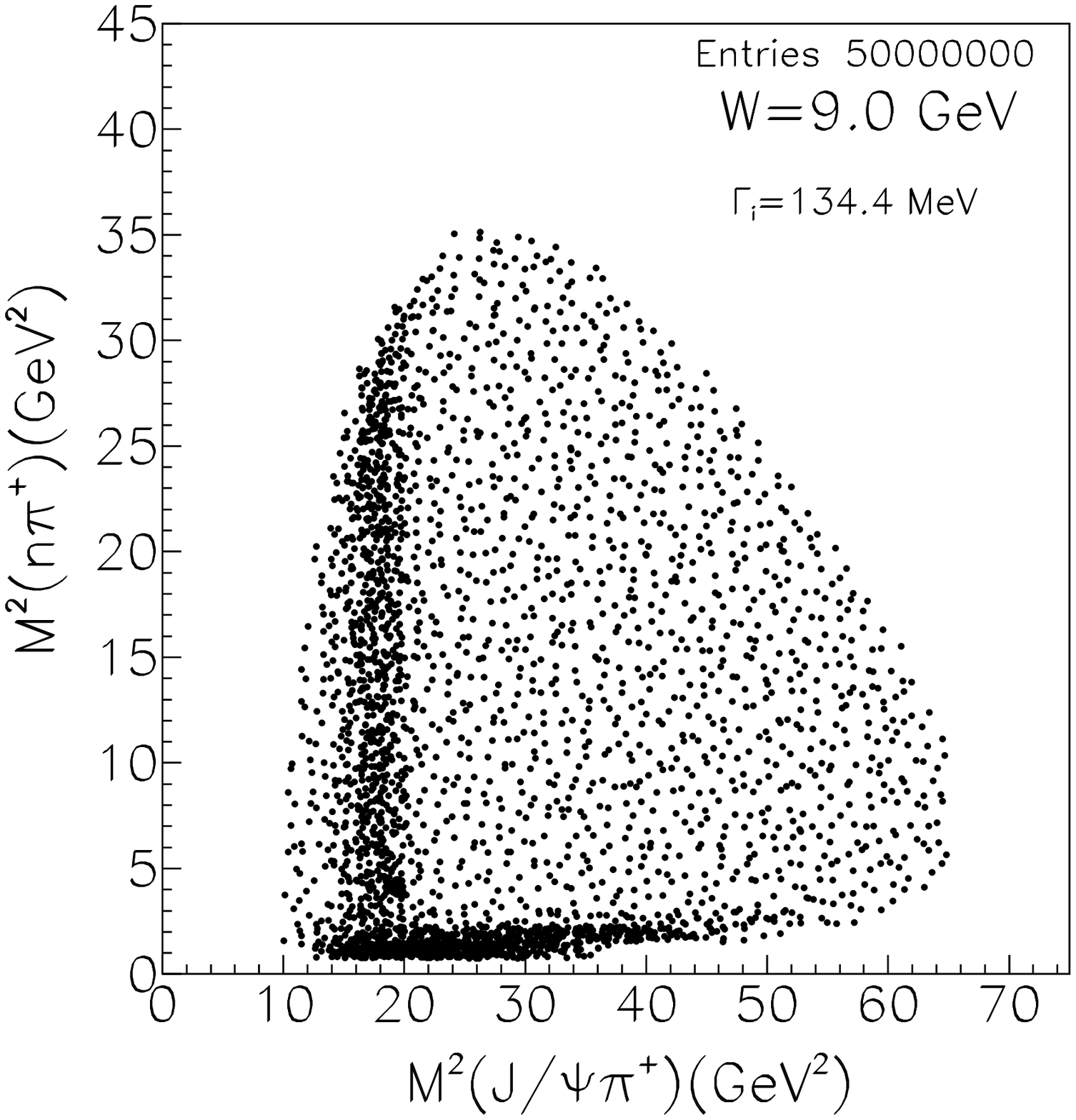}
\includegraphics[scale=0.29]{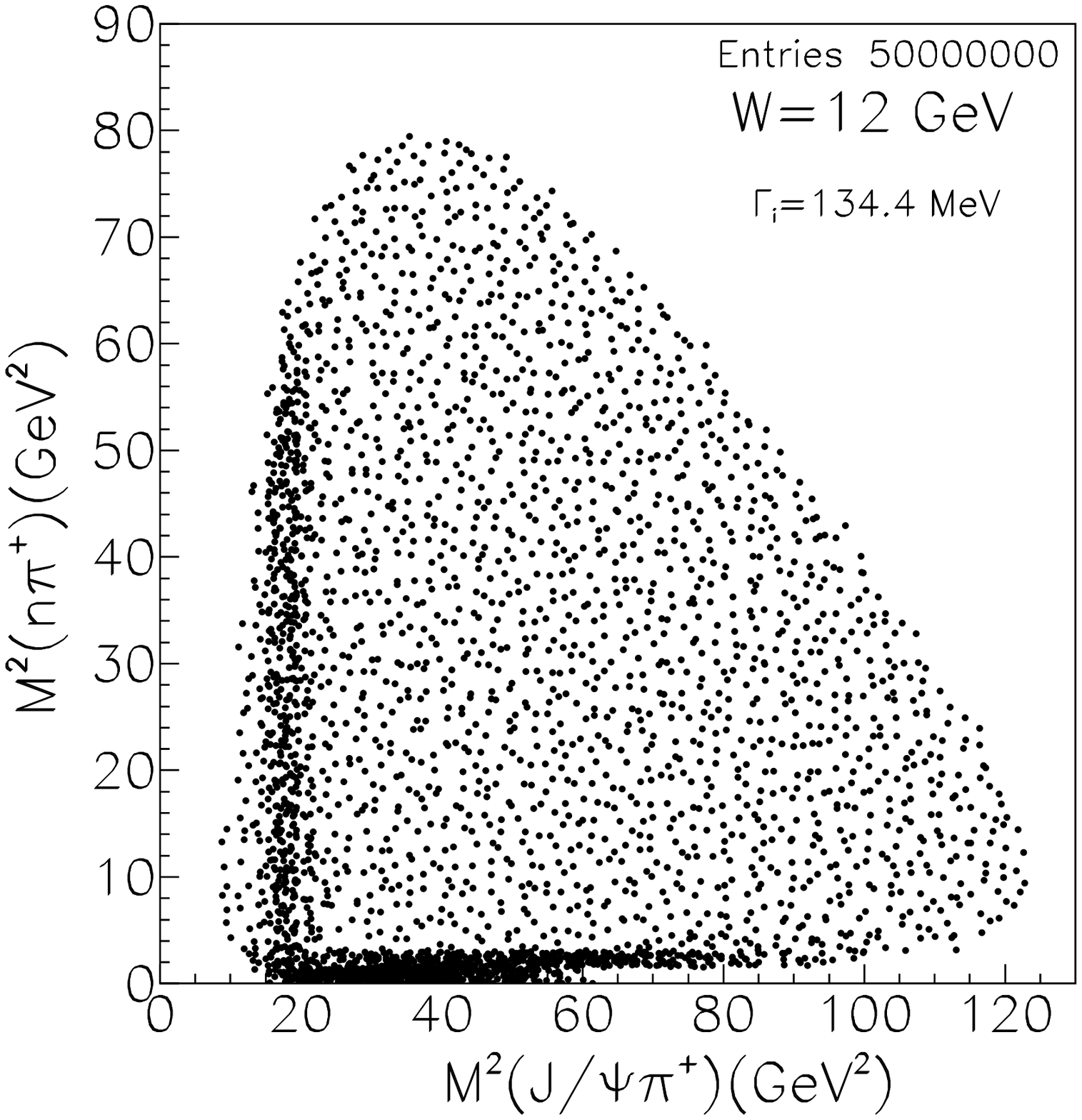}
\end{minipage}
\begin{minipage}{1\textwidth}
\includegraphics[scale=0.29]{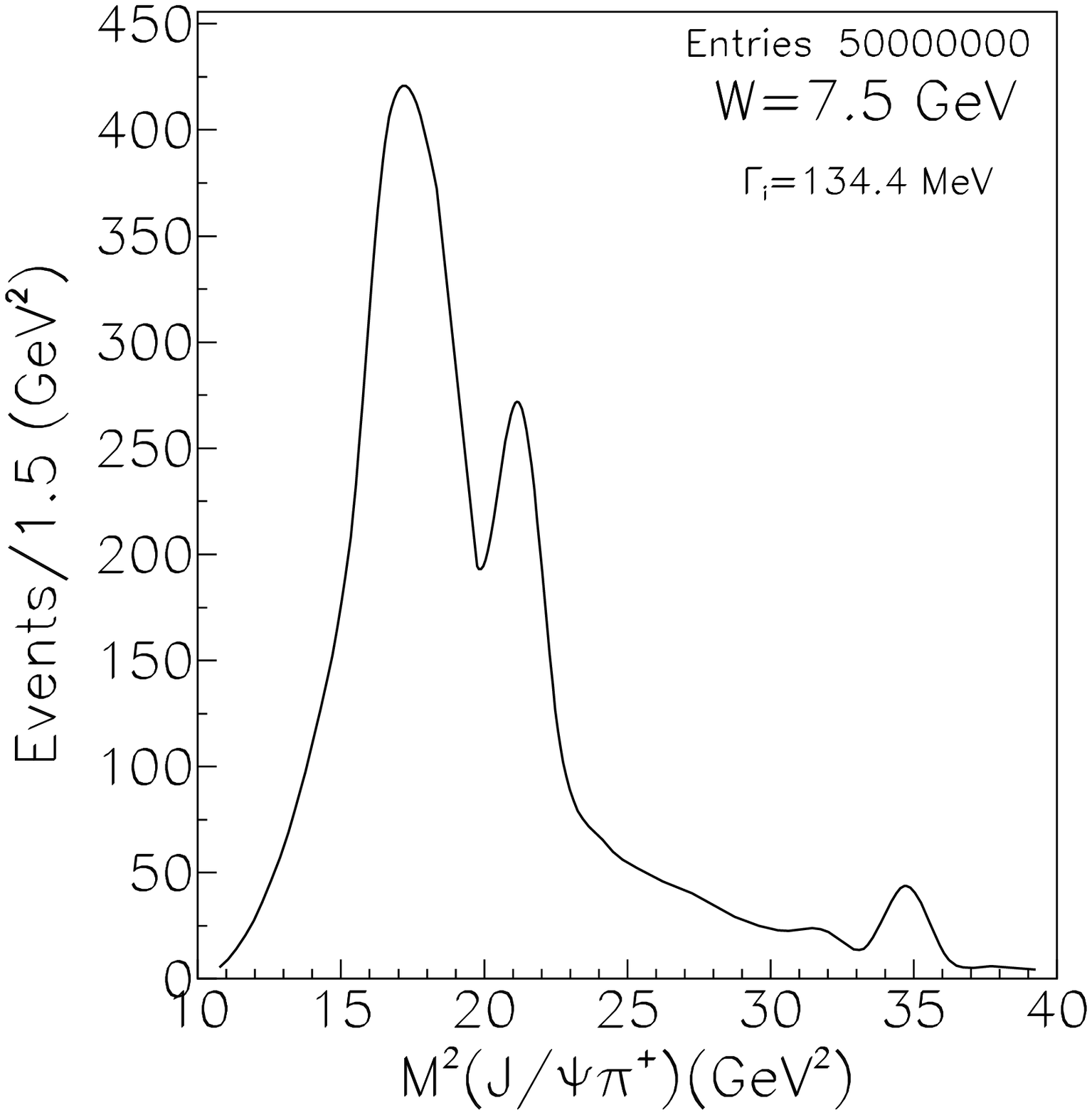}
\includegraphics[scale=0.29]{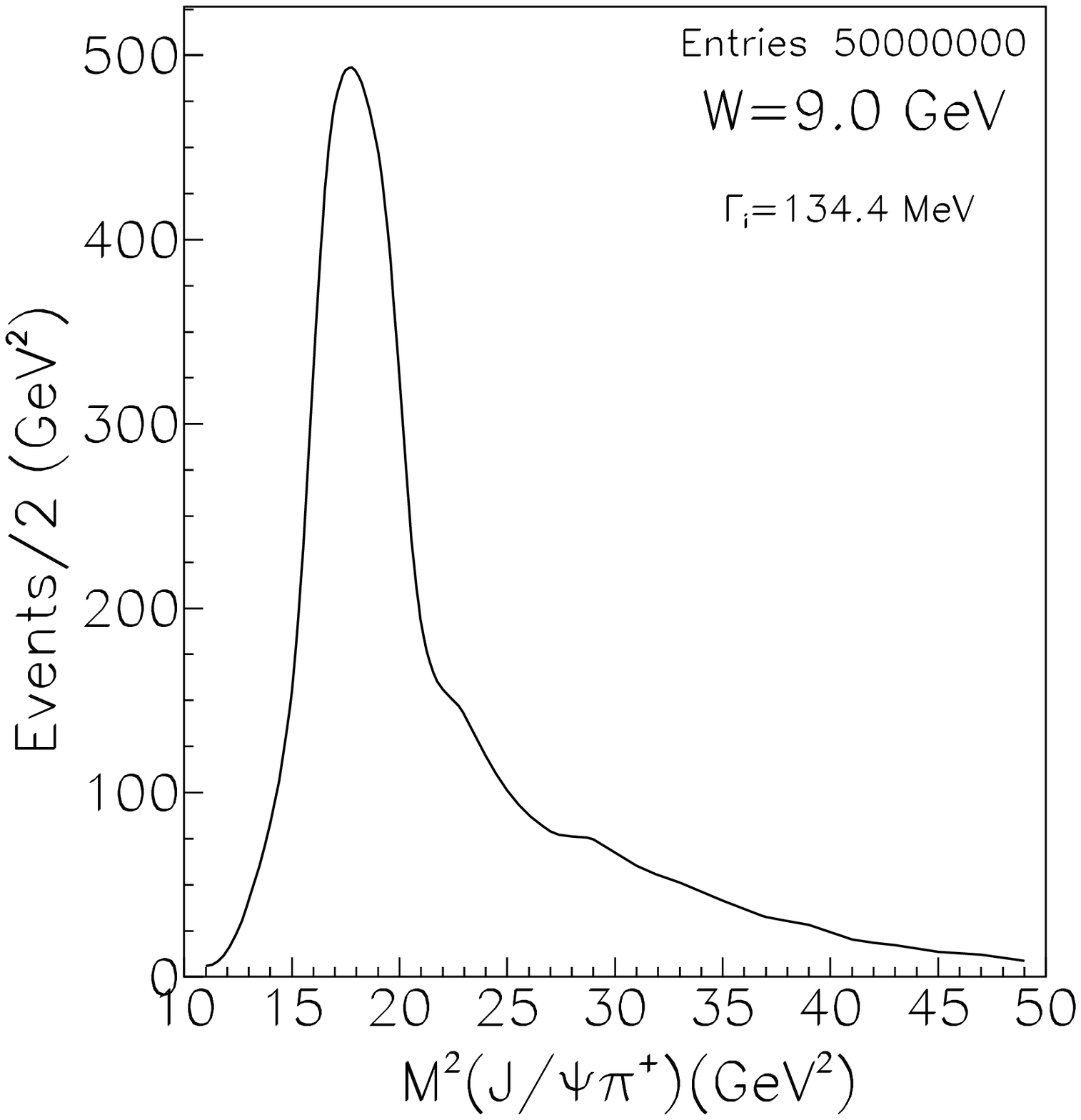}
\includegraphics[scale=0.29]{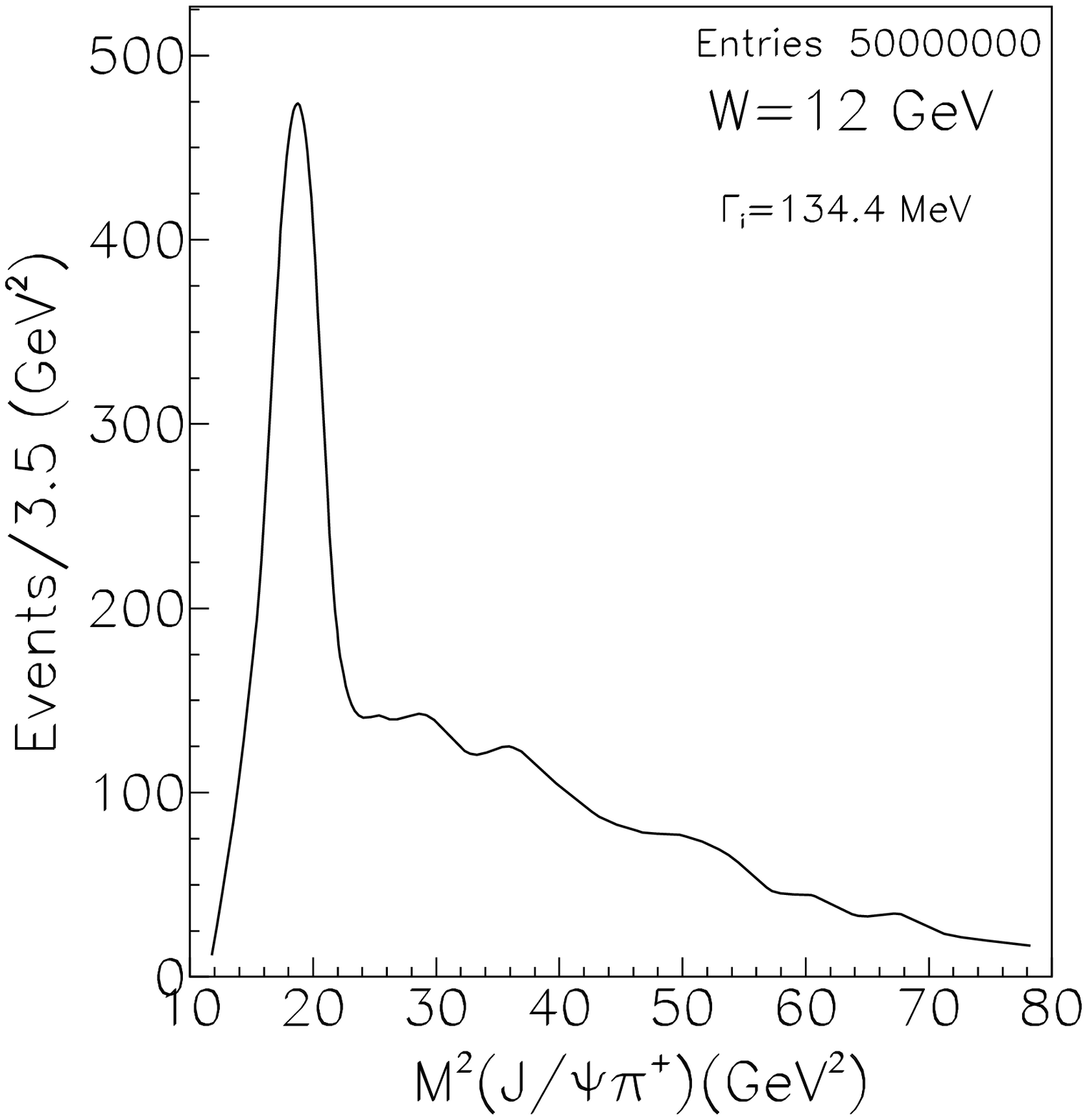}
\end{minipage}
\caption{(Color online) The Dalitz plot (top) and the $J/\protect\psi
\protect\pi ^{+}$ invariant mass spectrum (bottom) for the $\protect\gamma %
p\rightarrow J/\protect\psi \protect\pi ^{+}n$ reaction with the Reggeized
treatment at different center of mass energy $W=7.5,9,12$ GeV. Here, the
numerical result corresponds to the partial decay width $\Gamma
_{Z(4200)\rightarrow J/\protect\psi \protect\pi }=134.4$ MeV. }
\end{figure*}

To better understand that the effects of Reggeized treatment on the final
results, we calculate the total cross section of the $\gamma p\rightarrow
J/\psi \pi ^{+}n$ reaction without or with Reggeized treatment as presented
in Fig. 6 and Fig. 7, respectively.

Fig. 6 show the total cross sections for $\gamma p\rightarrow J/\psi \pi
^{+}n$ reaction including both $\pi $ exchange and Pomeron exchange
contributions by taking $\Lambda _{Z}=0.7$ GeV and $\Lambda _{N}=0.96$ GeV.
We notice that the line shape of total cross section goes up very rapidly
and has a peak around $W\simeq 7.5$ GeV. In this energy region, the cross
section of signal are larger than that of background when taking the partial
width values $\Gamma _{Z_{c}(4200)\rightarrow J/\psi \pi }=87.3$ or $134.4$
MeV.

In contrast, Fig. 7 present the total cross sections for $\gamma
p\rightarrow J/\psi \pi ^{+}n$ reaction including both pionic Regge
trajectory exchange and Pomeron exchange contributions by taking $\Lambda
_{Z}=0.7$ GeV and $\Lambda _{N}=0.96$ GeV. It is found that the total cross
section shows a peak at center of mass energy $W\simeq 9$ GeV, and the
contributions from signal are driven down when using the Reggeized
treatment. We note that the cross section of signal just a little bit higher
than that of background at center of mass energy $W\simeq 9$ GeV even a
larger partial decay width value ($\Gamma _{Z_{c}(4200)\rightarrow J/\psi
\pi }=134.4$ MeV) are adopted.

To demonstrate the feasibility of searching for the charged charmoniumlike $%
Z_{c}^{+}(4200)$ through the $\gamma p\rightarrow J/\psi \pi ^{+}n$
reaction, we further give the Dalitz plot and invariant mass spectrum for
the $\gamma p\rightarrow J/\psi \pi ^{+}n$ process.

Fig. 8 present the Dalitz plot and $J/\psi \pi ^{+}$ invariant mass spectrum
for the $\gamma p\rightarrow J/\psi \pi ^{+}n$ process with the Reggeized
treatment at different center of mass energy, where the numerical results
are obtained by taking the partial decay width $\Gamma
_{Z_{c}(4200)\rightarrow J/\psi \pi }=134.4$ MeV. From Dalitz plot we notice
that there exist a vertical band and a horizontal band, which are from the
signal and background contributions, respectively. Moreover, one notice that
the signal of $Z_{c}^{+}(4200)$ with $W=9.0$ is more explicit than that with
$W=7.5$ or 12 GeV, which is consistent with the result in Fig. 7. Thus we
can conclude that the $W=9.0$ GeV is the best energy window for searching
for the charged $Z_{c}^{+}(4200)$ via the $\gamma p\rightarrow J/\psi \pi
^{+}n$ process. By analyzing the $J/\psi \pi ^{+}$ invariant mass spectrum
in Fig. 8, one finds that the number of events of $J/\psi \pi ^{+}$ can
reach up to 500$/$2 GeV$^{2}$ at $W=9.0$ GeV when taking 50 million
collisions of $\gamma p$.
\begin{figure*}[t]
\begin{minipage}{1\textwidth}
\includegraphics[scale=0.35]{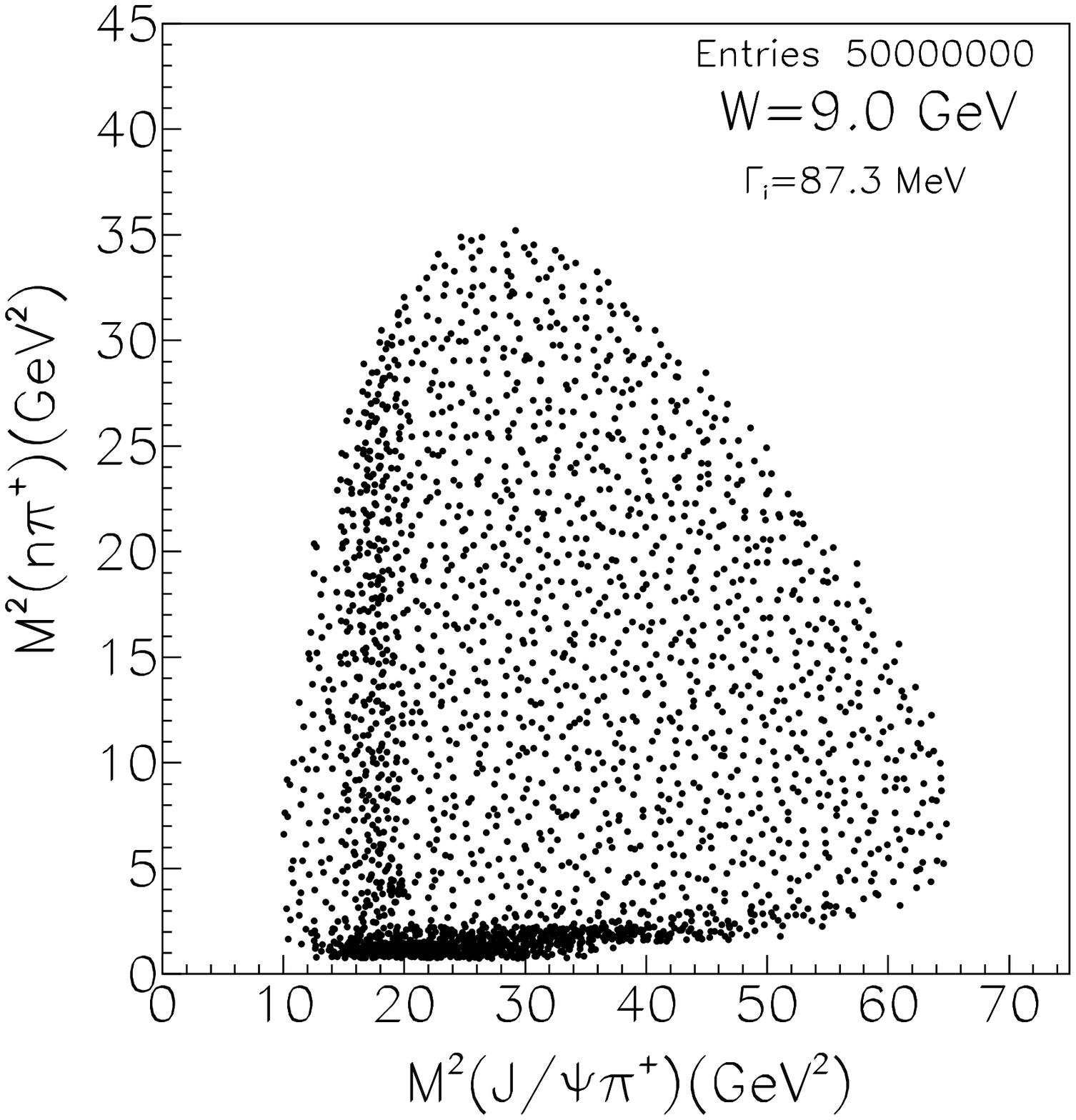}
\includegraphics[scale=0.35]{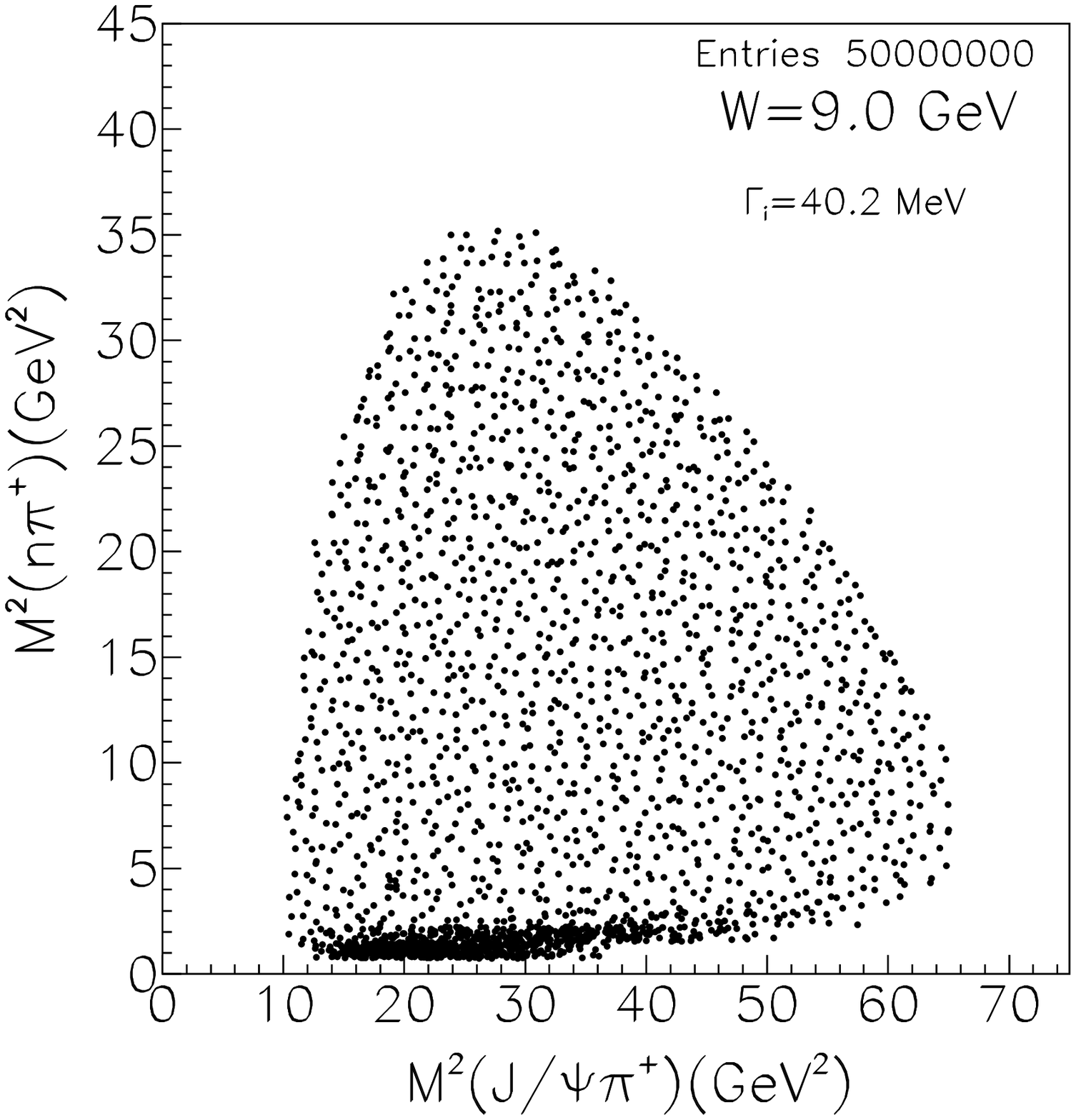}
\end{minipage}
\begin{minipage}{1\textwidth}
\includegraphics[scale=0.35]{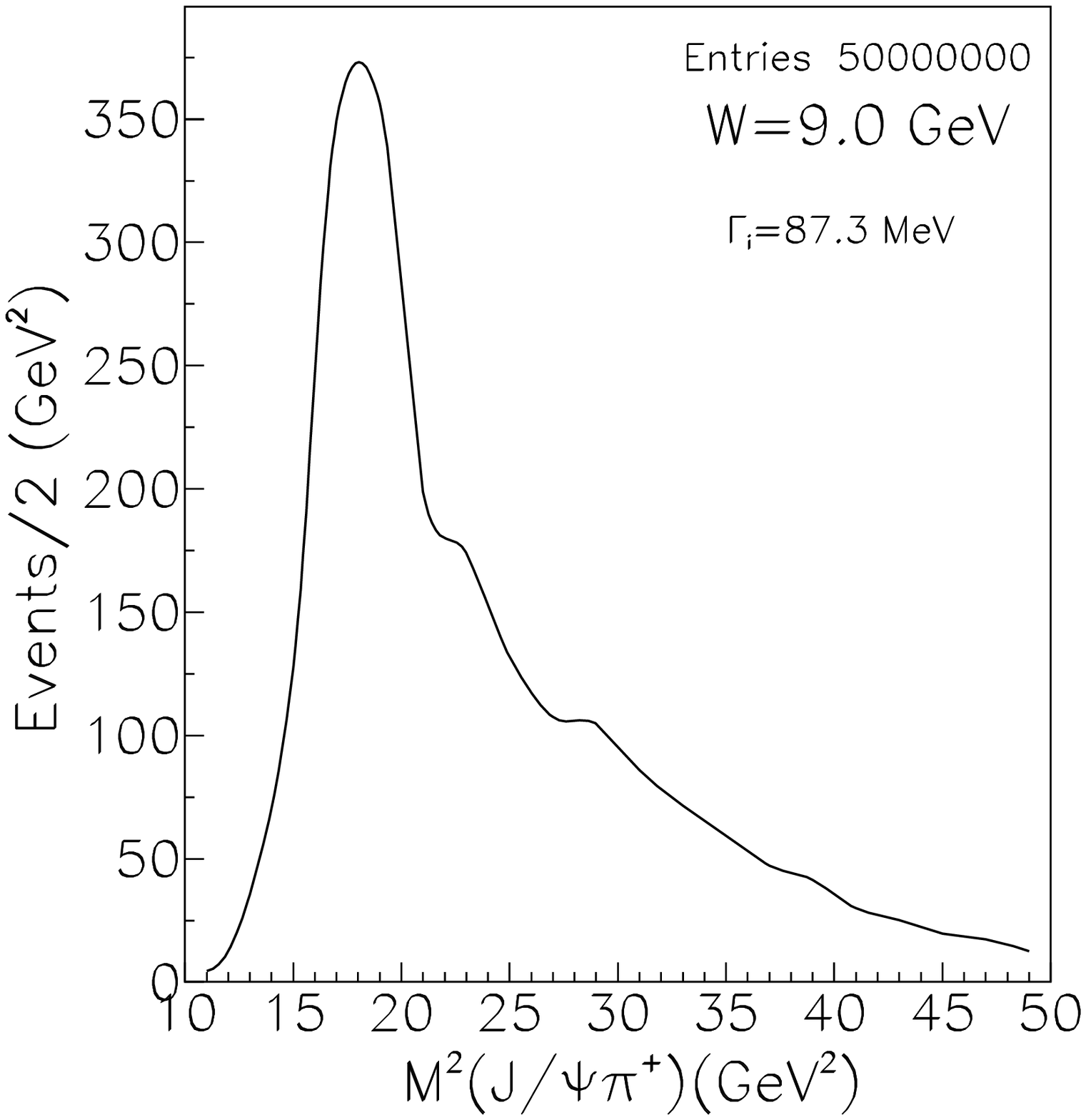}
\includegraphics[scale=0.35]{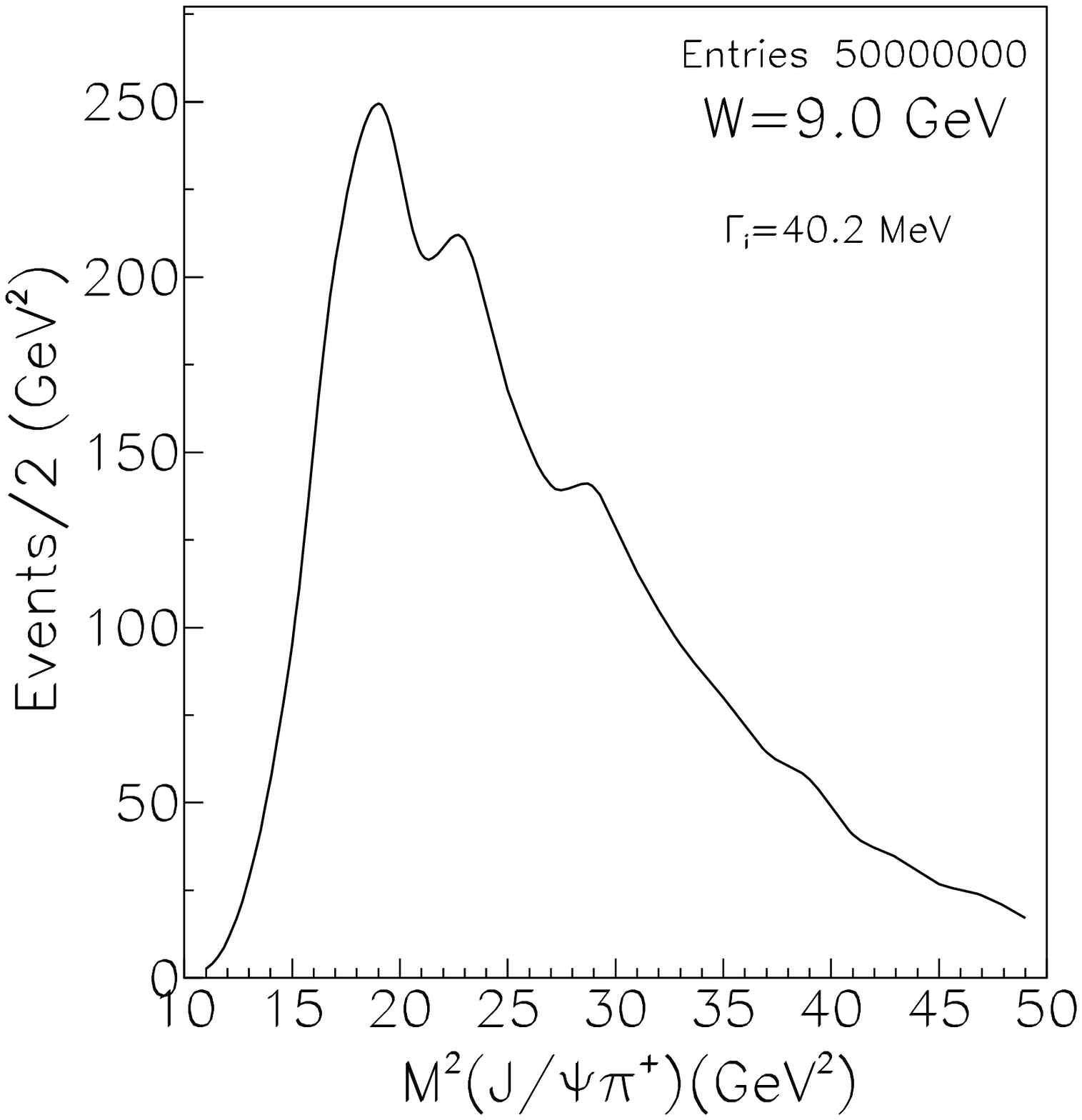}
\end{minipage}
\caption{(Color online) The Dalitz plot (top) and the $J/\protect\psi
\protect\pi ^{+}$ invariant mass spectrum (bottom) for the $\protect\gamma %
p\rightarrow J/\protect\psi \protect\pi ^{+}n$ reaction with the Reggeized
treatment at center of mass energy $W=9$ GeV. Here, the numerical results
correspond to the partial decay width values $\Gamma _{Z(4200)\rightarrow J/%
\protect\psi \protect\pi }=87.3,40.2$ MeV.}
\end{figure*}

Moreover, we take the center of mass energy $W=9.0$ GeV as one of the inputs
to calculate the Dalitz plot and $J/\psi \pi ^{+}$ invariant mass spectrum
related to the smaller partial decay width, which are presented in Fig. 9.
From Dalitz plot in Fig. 9 we notice that there exist an clear vertical band
which related to the $Z_{c}^{+}(4200)$ signal when taking partial decay
width $\Gamma _{Z_{c}(4200)\rightarrow J/\psi \pi }=87.3$ MeV. Since the
signal and background contribution do not interfere with each other as shown
in Dalitz plot, the signal of $Z_{c}^{+}(4200)$ can also be distinguished
from the background. Thus we can expect about 375$/$2 GeV$^{2}$ events for
the production of $J/\psi \pi ^{+}$ in 50 million collisions of $\gamma p$
at $W=9.0$ GeV if taking $\Gamma _{Z_{c}(4200)\rightarrow J/\psi \pi }=87.3$
MeV, which is enough to meet the requirements of the experiment. However, we
also see that the signal of $Z_{c}^{+}(4200)$ are submerged in the
background and will be difficult to distinguish it from the background if
taking $\Gamma _{Z_{c}(4200)\rightarrow J/\psi \pi }=40.2$ MeV.
\begin{figure*}[t]
\begin{minipage}{1\textwidth}
\includegraphics[scale=0.29]{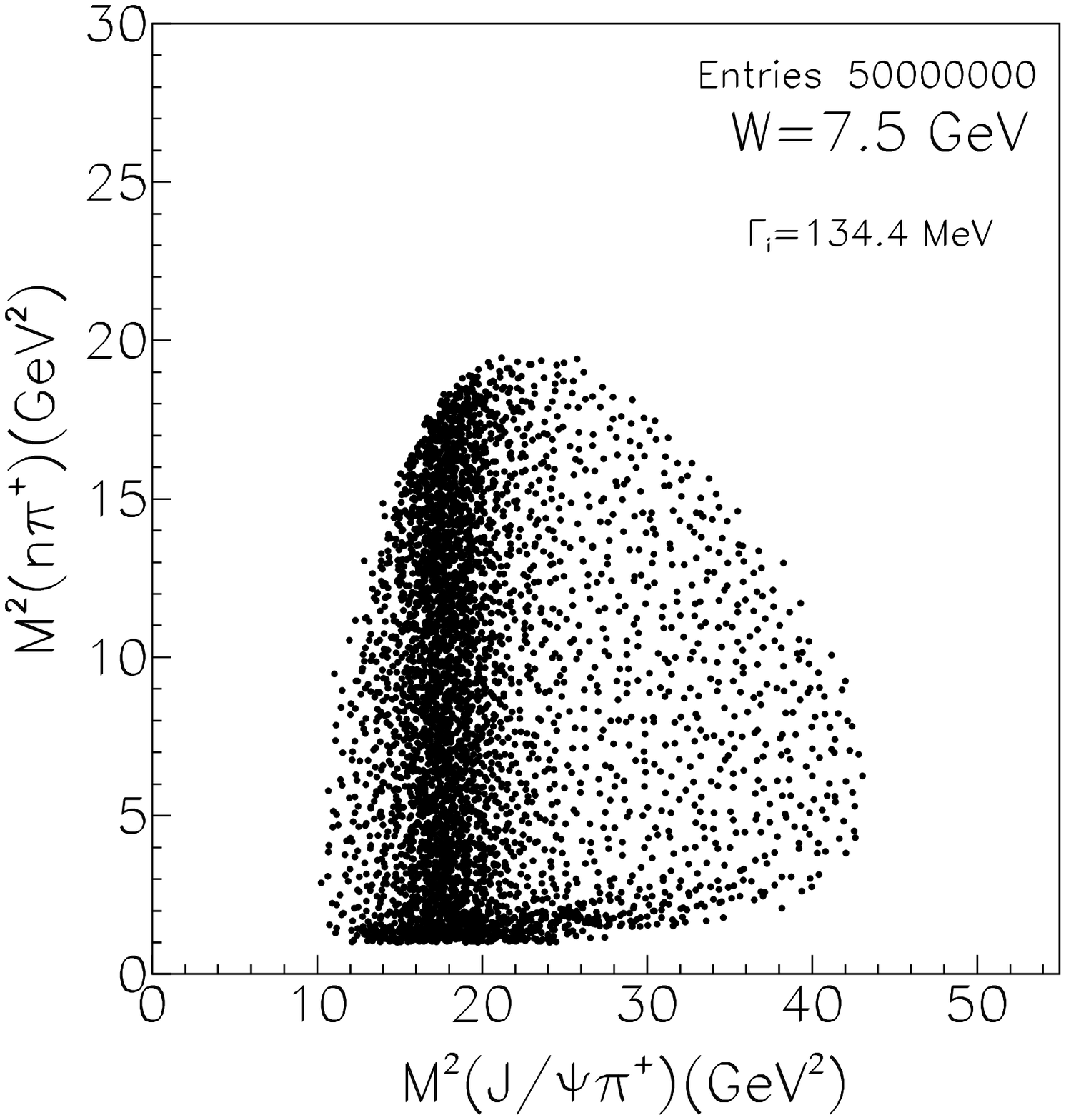}
\includegraphics[scale=0.29]{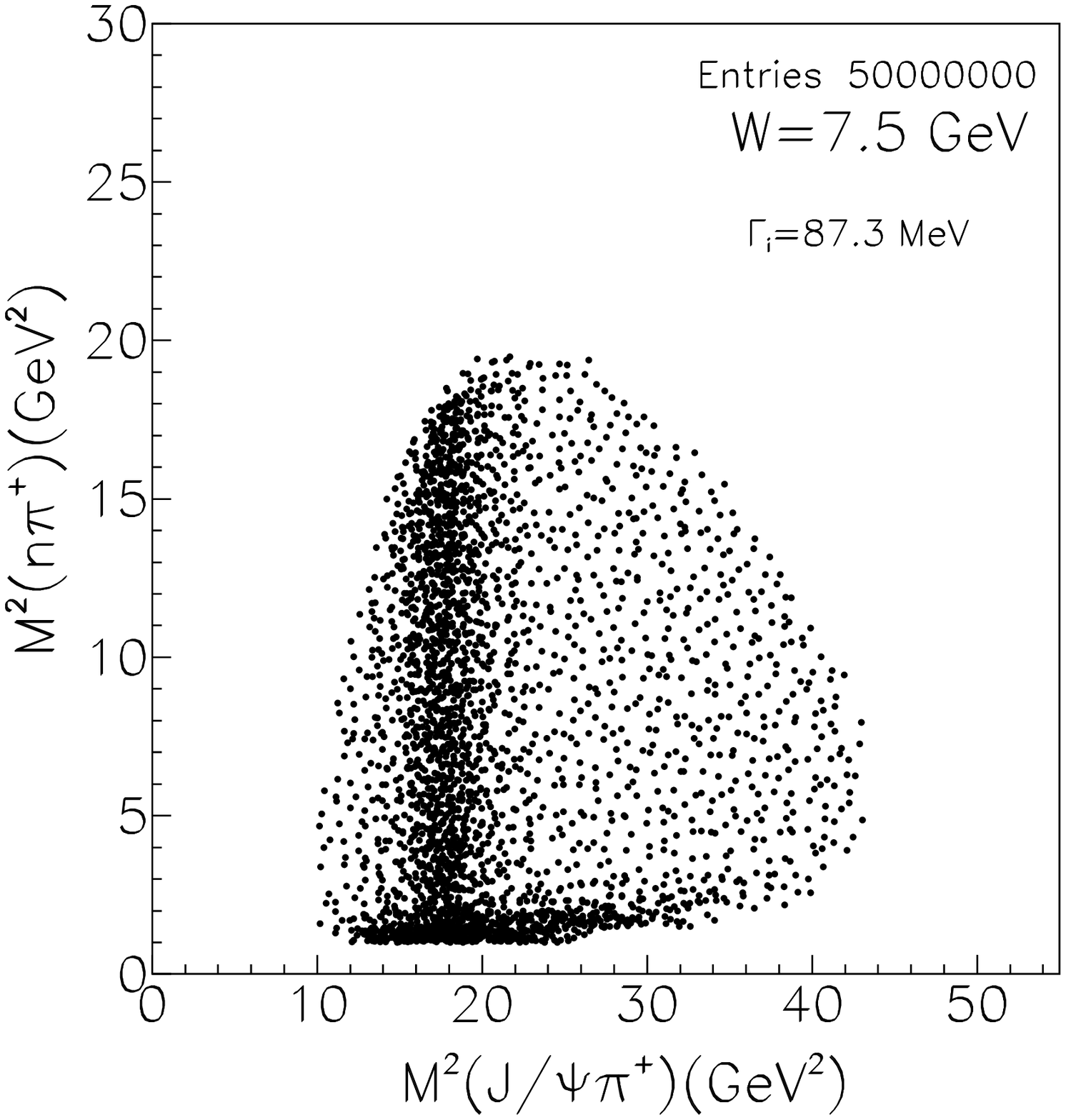}
\includegraphics[scale=0.29]{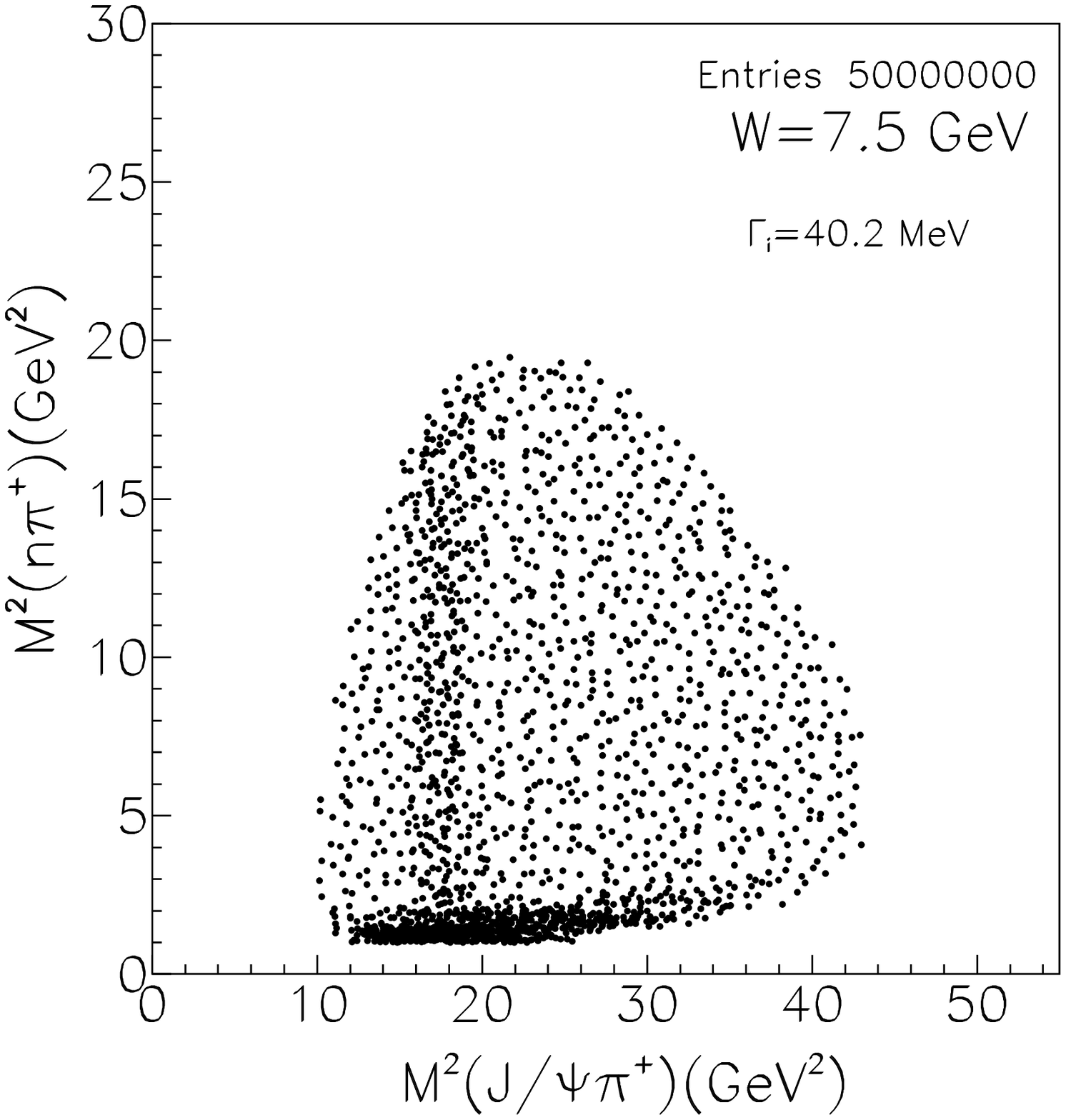}
\end{minipage}
\begin{minipage}{1\textwidth}
\includegraphics[scale=0.29]{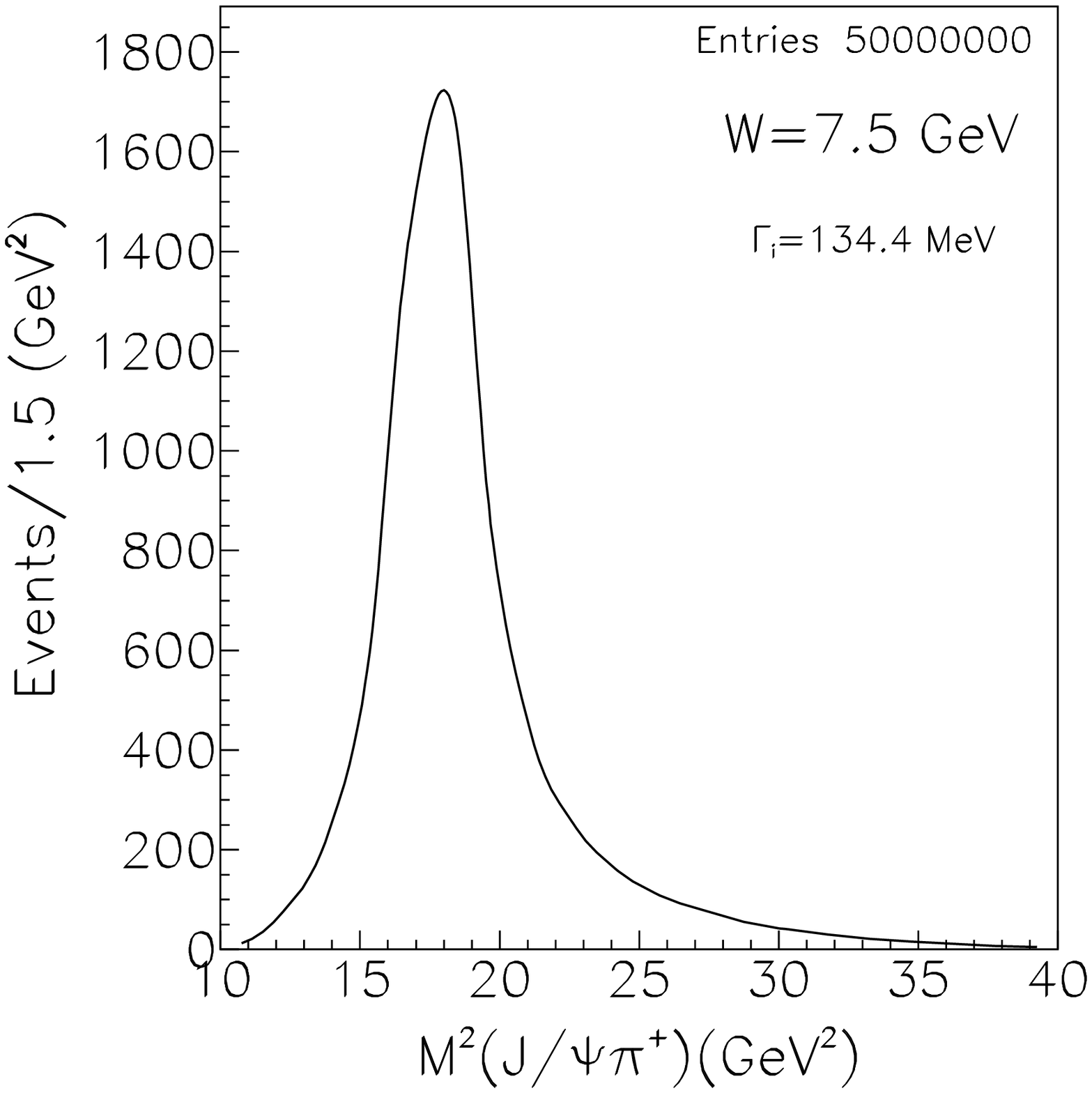}
\includegraphics[scale=0.29]{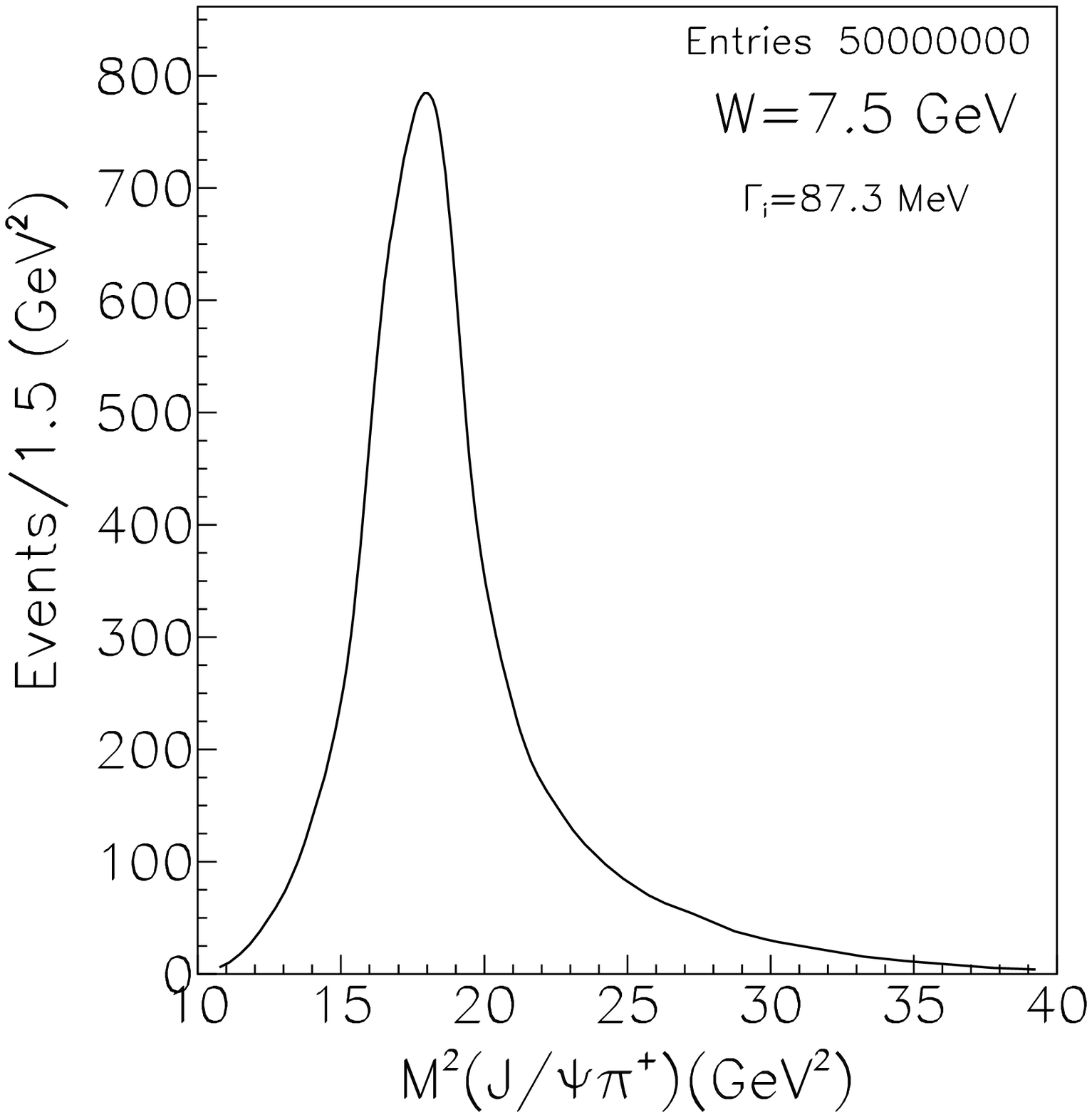}
\includegraphics[scale=0.29]{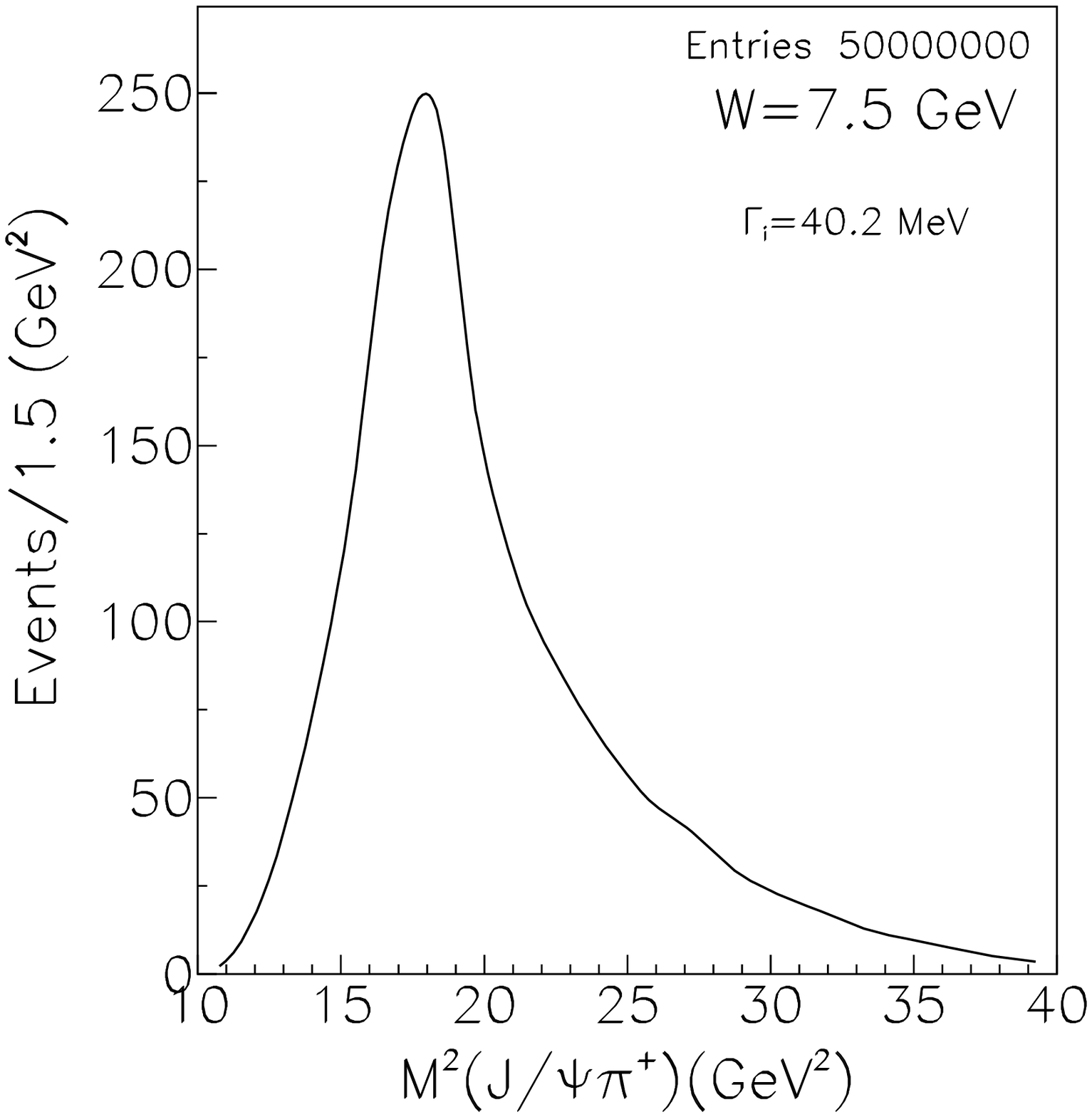}
\end{minipage}
\caption{(Color online) The Dalitz plot (top) and the $J/\protect\psi
\protect\pi ^{+}$ invariant mass spectrum (bottom) for the $\protect\gamma %
p\rightarrow J/\protect\psi \protect\pi ^{+}n$ reaction without the
Reggeized treatment at center of mass energy $W=7.5$ GeV. Here, the
numerical results correspond to the partial decay width values $\Gamma
_{Z(4200)\rightarrow J/\protect\psi \protect\pi }=134.4,87.3,40.2$ MeV.}
\end{figure*}

For comparison, we calculate the Dalitz plot and $J/\psi \pi ^{+}$ invariant
mass spectrum for the $\gamma p\rightarrow J/\psi \pi ^{+}n$ process without
the Reggeized treatment at $W=7.5$ GeV, as presented in Fig. 10. One finds
that a vertical band related to the signal of $Z_{c}^{+}(4200)$ appears in
Dalitz plot even if the lowest partial decay width ($\Gamma
_{Z_{c}(4200)\rightarrow J/\psi \pi }=40.2$ MeV) is adopted, which is
obvious different from that with Reggeized treatment.

\section{Upper limit of the decay width $\Gamma _{Z_{c}(4200)\rightarrow J/%
\protect\psi \protect\pi }$}

{The $J/\psi \pi ^{\pm }$ mass spectrum presented by the COMPASS
collaboration in \cite{compass}, which studied exclusive photoproduction of
a $J/\psi \pi ^{\pm }$ state at a nuclear target in the range from 7 GeV to
19 GeV in the centre-of-mass energy of the photon-nucleon system, does not
exhibit any statistically significant structure at about $4.2$ GeV.
Nevertheless it can be used for estimation of an upper limit for the value $%
BR(Z_{c}(4200)\rightarrow J/\psi \pi )\times \sigma _{\gamma N\rightarrow
Z_{c}(4200)N}$. }

A sum of two exponential functions for a continuum and a Breit-Wigner curve
for a possible contribution of $Z_{c}^{\pm }(4200)$ photoproduction was
fitted to the mass spectrum in the range from 3.4 GeV to 6.0 GeV. The mass $%
M_{Z_{c}(4200)}=4196$~MeV and the width $\Gamma _{Z_{c}(4200)}=370$~MeV were
used as the fixed parameters. Doing this we ignore possible contribution of
any other resonances like $Z_{c}(3900)$ and their interference with $%
Z_{c}(4200)$. The $J/\psi \pi ^{\pm }$ mass spectrum with the fitting curve
is shown in Fig. 11. The obtained from the fit possible number of $%
Z_{c}(4200)$ events is $N_{Z_{c}(4200)}=58\pm 31$. It can be converted to
the upper limit $N_{Z_{c}(4200)}^{UL}<98$ events corresponding to a
confidence level of CL = 90\%. According to the normalization used in \cite%
{compass} this limit corresponds to the result
\begin{equation}
BR(Z_{c}(4200)\rightarrow J/\psi \pi )\times \sigma _{\gamma N\rightarrow
Z_{c}(4200)N}<340~\mathrm{pb}.
\end{equation}

This result can be used for estimation of an upper limit for the partial
width $\Gamma _{J/\psi \pi }$ of the decay $Z_{c}({4200)}\rightarrow J/\psi
\pi $ based on the Reggeized treatment. The production cross section,
averaged over the $W$-range covered by COMPASS, is about $\Gamma _{J/\psi
\pi }\times 91$~pb/MeV. So
\begin{equation}
\frac{\Gamma _{J/\psi \pi }}{\Gamma _{tot}}\times \sigma _{\gamma
N\rightarrow Z_{c}^{\pm }(4200)~N}=\frac{\Gamma _{J/\psi \pi }^{2}\times 90~%
\mathrm{pb/MeV}}{\Gamma _{tot}}<340~\mathrm{pb}.
\end{equation}%
Assuming $\Gamma _{\mbox{tot}}=370$~MeV, we obtain an upper limit $\Gamma
_{J/\psi \pi }<37$ MeV.

Photoproduction of the $Z_{c}^{+}(4200)$ state could also be tested using
the data on the HERMES experiment. It covers the range $2$ GeV$<W<6.3$ GeV
\cite{HERMES} where the difference between production cross sections
calculated through pionic Regge trajectory exchange and virtual pion
exchange is maximal.
\begin{figure}[tb]
\centering
\includegraphics[scale=0.4]{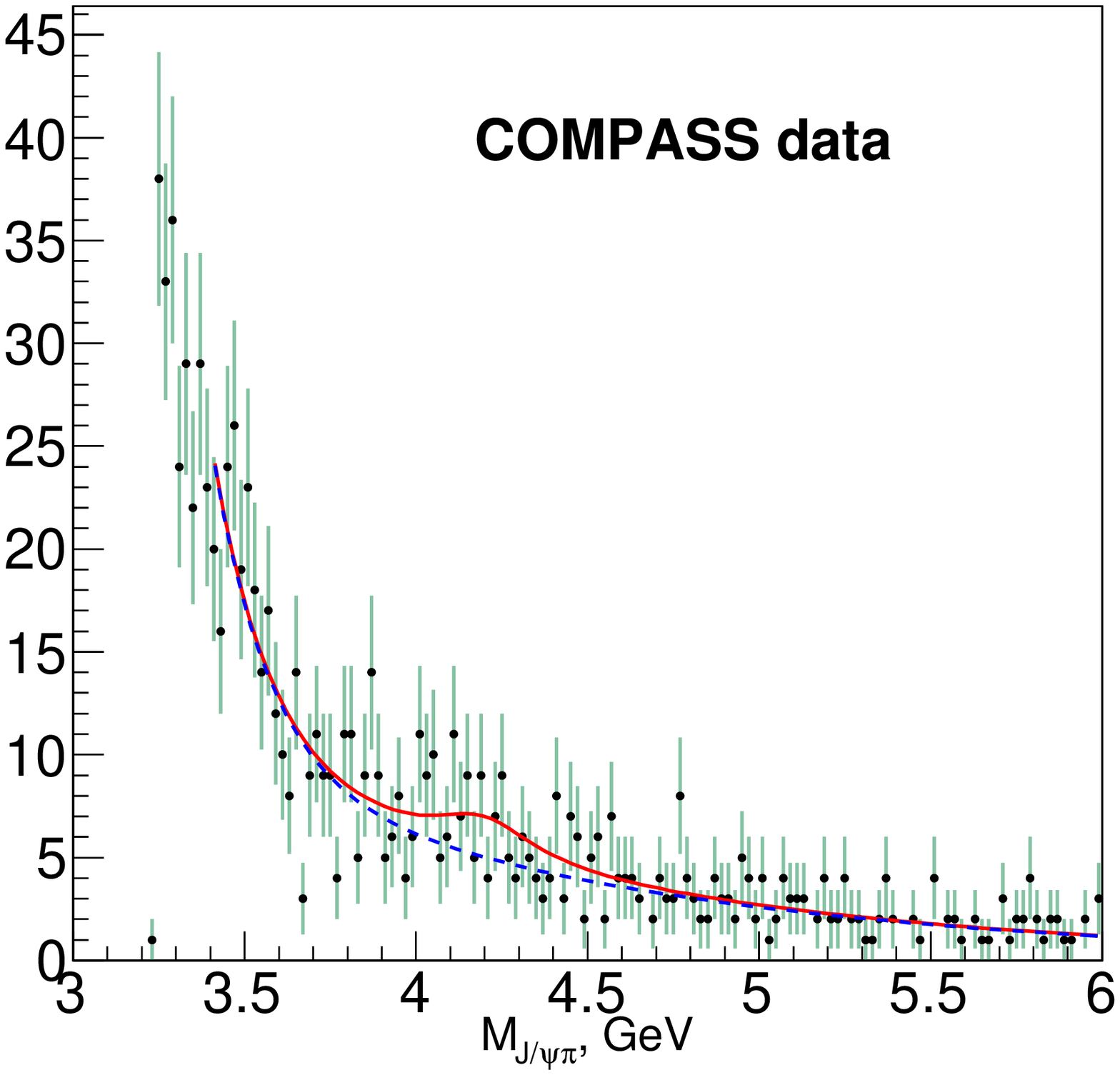}
\caption{(Color online) Mass spectrum of the $J/\protect\psi \protect\pi $
state obtained by COMPASS \protect\cite{compass}. The fitted function is
shown as a red solid line. Dashed blue line corresponds to the continuum
description.}
\end{figure}

\section{Summary}

In this work, we study the charged $Z_{c}(4200)$ production in $\gamma
p\rightarrow J/\psi \pi ^{+}n$ reaction with an effective Lagrangian
approach and the Regge trajectories model. Since the charmoniumlike $%
Z_{c}(4200)$ was only observed in $B$ meson decay process, it is an
interesting and important topic to study the charmoniumlike $Z_{c}(4200)$ by
different processes.

Through of analysis and comparison, our numerical results indicate:

\begin{itemize}
\item[(I)] The effect of introducing the Reggeized treatment has been to
significantly reduce the magnitude of cross section for the $Z_{c}(4200)$
photoproduction. The total cross section for the $\gamma p\rightarrow
Z_{c}^{+}(4200)n$ via pionic Regge trajectory exchange is smaller than that
of without Reggeized treatment and the predictions in Refs. \cite%
{lxh08,he09,lin13,lin14}.

\item[(II)] We finds that the differential cross section for the $\gamma
p\rightarrow Z_{c}^{+}(4200)n$ by exchanging the pionic Regge trajectory are
very sensitive to the $\theta $ angle and gives a considerable contribution
at forward angles, which can be checked by further experiment and may be an
effective way to examine the validity of the Reggeized treatment.

\item[(III)] The total cross section for the $\gamma p\rightarrow J/\psi \pi
^{+}n$ process with Reggeized treatment is lower than that of without
Reggeized treatment. The calculations indicate that the partial decay width $%
\Gamma _{Z_{c}(4200)\rightarrow J/\psi \pi }$ is a key parameter in studying
the production of $Z_{c}(4200)$ via $\gamma p$ collision. Adopting the
partial decay width predicted in Ref. \cite{wei15} by assuming that the $%
Z_{c}(4200)$ is a tetraquark state, we find that the signal of $%
Z_{c}^{+}(4200)$ can also be distinguished from the background at $W=9.0$
GeV if taking $\Gamma _{Z_{c}(4200)\rightarrow J/\psi \pi }=87.3$ MeV, but
not for the case of taking $\Gamma _{Z_{c}(4200)\rightarrow J/\psi \pi
}=40.2 $ MeV. In Ref. \cite{wzg15}, by assuming the $Z_{c}(4200)$ as an
axial-vector molecule-like state, the partial decay width $\Gamma
_{Z_{c}(4200)\rightarrow J/\psi \pi }=24.6$ MeV was obtained with QCD sum
rule. If the predicted $\Gamma _{Z_{c}(4200)\rightarrow J/\psi \pi }=24.6$
MeV in Ref. \cite{wzg15} is reliable, then the signal of $Z_{c}^{+}(4200)$
produced in $\gamma p$ collision will be difficult to be distinguished from
background. Thus the experiment of the meson photoproduction of $Z_{c}(4200)$
may provide a useful adjunctive information for the confirmation of the
inner structure of $Z_{c}(4200)$.

\item[(IV)] The peak position of total cross section for the $\gamma
p\rightarrow Z_{c}^{+}(4200)n\rightarrow J/\psi \pi ^{+}n$ process was moved
to the higher energy point when adding the Reggeized treatment, which means
that a higher beam energy is necessary for the meson photoproduction of $%
Z_{c}(4200)$. The results show that $W\simeq 9.0$ GeV is the best energy
window for searching for the $Z_{c}(4200)$ via $\gamma p$ collision. All
these calculations can be checked in the future experiment.

\item[(V)] Using data on exclusive photoproduction of a $J/\psi \pi ^{\pm }$
state from COMPASS we estimated the upper limit for the value of $%
Z_{c}(4200) $ production cross section multiplied by the branching ratio of
the $Z_{c}(4200)\rightarrow J/\psi \pi $ decay to be above $340$~pb (CL=90
\%). According to the Reggeized treatment it corresponds to the upper limit
of $\Gamma _{Z_{c}(4200)\rightarrow J/\psi \pi }$ of about 37 MeV, which is
coincidence with the prediction of $\Gamma _{Z_{c}(4200)\rightarrow J/\psi
\pi }=24.6$ MeV by assuming the $Z_{c}(4200)$ as a molecule-like state in
\cite{wzg15}.
\end{itemize}

Since the Reggeized treatment used in this work has been proven to be more
precise than the general effective Lagrangian approach in the pion and kaon
photoproduction \cite{mg97,gg11,he14}, our theoretical results may provide a
valuable information, both for searching for the $Z_{c}(4200)$ via $\gamma p$
collision or explaining the lack of observation of $Z_{c}(4200)$ in
experiment. Therefore, the more experiment about the photoproduction of $%
Z_{c}(4200)$ are suggested, which will be important to improve our knowledge
of the nature of $Z_{c}(4200)$ and the Regge theory.

\section{Acknowledgments}

The authors would like to acknowledge the COMPASS collaboration for allowing
us to use the {data of $J/\psi \pi ^{\pm }$ mass spectrum}. Meanwhile, X. Y.
W. is grateful Dr. Qing-Yong Lin for the valuable discussions and help. This
work is partly supported by the National Basic Research Program (973 Program
Grant No. 2014CB845406), the National Natural Science Foundation of China
(Grant No. 11175220) and the One Hundred Person Project of the Chinese
Academy of Science (Grant No. Y101020BR0).


\begin{thebibliography}{99}
\bibitem{belle03} S. K. Choi \textit{et al.} (Belle Collaboration), Phys.
Rev. Lett. 91, 262001 (2003).

\bibitem{lhc13} R. Aaij \textit{et al.} (LHCb Collaboration), Phys. Rev.
Lett. 110, 222001 (2013).

\bibitem{ba05} B. Aubert \textit{et al.} (BARBAR Collaboration), Phys. Rev.
Lett. 95, 142001 (2005).

\bibitem{belle07} C. Z. Yuan \textit{et al.} (Belle Collaboration), Phys.
Rev. Lett. 99, 182004 (2007).

\bibitem{bes13} M. Ablikim \textit{et al.} (BESIII Collaboration), Phys.
Rev. Lett. 110, 252001 (2013).

\bibitem{belle110} Z. Q. Liu \textit{et al.} (Belle Collaboration), Phys.
Rev. Lett. 110, 252002 (2013).

\bibitem{bes14} M. Ablikim \textit{et al.} (BESIII Collaboration), Phys.
Rev. Lett. 112, 132001 (2014).

\bibitem{bes111} M. Ablikim \textit{et al.} (BESIII Collaboration), Phys.
Rev. Lett. 111, 242001 (2013).

\bibitem{belle08} R. Mizuk \textit{et al.} (Belle Collaboration), Phys. Rev.
D 78, 072004 (2008).

\bibitem{belleprd} S. K. Choi \textit{et al.} (Belle Collaboration), Phys.
Rev. Lett. 100, 142001 (2008).

\bibitem{belle100} K. Chilikin \textit{et al.} (Belle Collaboration), Phys.
Rev. D 88, 074026 (2013).

\bibitem{lhc14} R. Aaij \textit{et al.} (LHCb Collaboration), Phys. Rev.
Lett. 112, 222002 (2014).

\bibitem{belle14} K. Chilikin \textit{et al.} (Belle Collaboration), Phys.
Rev. D 90, 112009 (2014).

\bibitem{ab12} A. Bondar \textit{et al.} (Belle Collaboration), Phys. Rev.
Lett. 108, 122001 (2012).

\bibitem{ia12} I. Adachi \textit{et al.} (Belle Collaboration), Phys. Rev.
Lett. 108, 032001 (2012).

\bibitem{iab} I. Adachi \textit{et al.} (Belle Collaboration),
arXiv:1105.4583.

\bibitem{lx14} X. Liu, Chin. Sci. Bull. 59, 3815 (2014).

\bibitem{ae14} A. Esposito, A. L. Guerrieri, F. Piccinini, A. Pilloni and A.
D. Polosa, Int. J. Mod. Phys. A 30, 1530002 (2014).

\bibitem{sl14} S. L. Olsen, Hyperfine Interact. 229, 7 (2014).

\bibitem{mn10} M. Nielsen, F. S. Navarra and S. H. Lee, Phys. Rept. 497, 41
(2010).

\bibitem{mn14} M. Nielsen and F. S. Navarra, Mod. Phys. Lett. A 29, 1430005
(2014).

\bibitem{tx13} T. Xiao \textit{et al.,} Phys. Lett. B 727, 366 (2013).

\bibitem{zb13} P. Krokovny \textit{et al.} (Belle Collaboration), Phys. Rev.
D 88, 052016 (2013).

\bibitem{zc14} M. Ablikim \textit{et al.} (BESIII Collaboration), Phys. Rev.
Lett. 113, 212002 (2014).

\bibitem{zsl05} S. L. Zhu, Phys. Lett. B 625, 212 (2005).

\bibitem{zsl10} Y.-R. Liu \textit{et al.,} Phys. Rev. D 82, 014011 (2010).

\bibitem{ln12} N. Li and S.-L. Zhu, Phys. Rev. D 86, 074022 (2012).

\bibitem{hh06} H. Hogaasen, J. M. Richard, and P. Sorba, Phys. Rev. D 73,
054013 (2006).

\bibitem{de06} D. Ebert, R. N. Faustov, and V. O. Galkin, Phys. Lett. B 634,
214 (2006).

\bibitem{nb06} N. Barnea, J. Vijande, and A. Valcarce, Phys. Rev. D 73,
054004 (2006).

\bibitem{qw13} Q. Wang, C. Hanhart, and Q. Zhao, Phys. Rev. Lett. 111,
132003 (2013).

\bibitem{eb13} E. Braaten, Phys. Rev. Lett. 111, 162003 (2013).

\bibitem{dy13} D. Y. Chen, X. Liu, and T. Matsuki, Phys. Rev. D 88, 036008
(2013).

\bibitem{cf14} C. F. Qiao and L. Tang, Eur. Phys. J. C 74, 2810 (2014).

\bibitem{fa14} F. Aceti, M. Bayar, and E. Oset, Eur. Phys. J. A 50, 103
(2014).

\bibitem{iv12} I. V. Danilkin, V. D. Orlovsky and Yu. A. Simonov, Phys. Rev.
D 85, 034012 (2012).

\bibitem{db11} D. Bugg, Europhys. Lett. 96, 11002 (2011).

\bibitem{rdm07} R. D. Matheus, \textit{et al.,} Phys. Rev. D 75, 014005
(2007).

\bibitem{lm14} L. Maiani \textit{et al.,} Phys. Rev. D 89, 114010 (2014).

\bibitem{zl14} L. Zhao, W. Z. Deng and S. L. Zhu, Phys. Rev. D 90, 094031
(2014).

\bibitem{wei11} W. Chen and S. L. Zhu, Phys. Rev. D 83, 034010 (2011).

\bibitem{wei12} W. Chen and S. L. Zhu, EPJ Web Conf. 20, 01003 (2012).

\bibitem{wei15} W. Chen \textit{et al., }arXiv:1501.03863 [hep-ph].

\bibitem{wjj10} J. J. Wu, R. Molina, E. Oset and B. S. Zou, Phys. Rev. Lett.
105, 232001 (2010).

\bibitem{wjj11} J. J. Wu, R. Molina, E. Oset and B. S. Zou, Phys. Rev. C 84,
015202 (2011).

\bibitem{xyw15} X. Y. Wang and X. R. Chen, Europhys. Lett. 109, 41001 (2015).

\bibitem{xy2015} X. Y. Wang and X. R. Chen, Eur. Phys. J. A 51, 85 (2015).

\bibitem{wzg15} Z. G. Wang, arXiv:1502.01459.

\bibitem{ke08} H. W. Ke and X. Liu, Eur. Phys. J. C 58, 217 (2008).

\bibitem{wang15} X. Y. Wang, J. J. Xie and X. R. Chen, Phys. Rev. D 91,
014032 (2015).

\bibitem{ahep15} X. Y. Wang, X. R. Chen, Adv. High Energy Phys. 2015, 918231
(2015).

\bibitem{lxh08} X. H. Liu, Q. Zhao, and F. E. Close, Phys. Rev. D 77,
0944005 (2008).

\bibitem{he09} J. He and X. Liu, Phys. Rev. D 80, 114007 (2009).

\bibitem{lin13} Q. Y. Lin \textit{et al.,} Phys. Rev. D 88, 114009 (2013).

\bibitem{lin14} Q. Y. Lin \textit{et al.,} Phys. Rev. D 89, 034016 (2014).

\bibitem{compass} C. Adolph \textit{et al.} (COMPASS Collaboration), Phys.
Lett. B 742, 330 (2015).

\bibitem{mg97} M. Guidal, J. M. Laget, and M. Vanderhaeghen, Nucl. Phys. A
627, 645 (1997).

\bibitem{gg11} G. Galat$\grave{a}$, Phys. Rev. C 83, 065203 (2011).

\bibitem{he14} J. He, Phys. Rev. C 89, 055204 (2014).

\bibitem{ab06} A. B. Kaidalov \textit{et al., }Eur. Phys. J. C 47, 385
(2006).

\bibitem{va06} V. A. Khoze, A. D. Martin and M. G. Ryskin, Eur. Phys. J. C
48, 797 (2006).

\bibitem{sc02} S. Chekanov, \textit{et al.} (ZEUS Collaboration), Nucl.
Phys. B 637, 3 (2002).

\bibitem{bz96} B. Z. Kopeliovich, B. Povh and I. Potashnikova, Z. Phys. C
73, 125 (1996).

\bibitem{hh95} H. Holtmann, \textit{et al., }Phys. Lett. B 338, 393 (1995).

\bibitem{mb13} M. Burkardt \textit{et al.,} Phys. Rev. D 87, 056009 (2013).

\bibitem{ys15} Y. Salamu \textit{et al., }Phys. Rev. Lett. 114, 122001
(2015).

\bibitem{fc} F. Carvalho \textit{et al., }arXiv:1507.07758.

\bibitem{kt94} K. Tsushima, S. W. Huang, and A. Faessler, Phys. Lett. B 337,
245 (1994).

\bibitem{kt97} K. Tsushima, \textit{et al.,} Phys. Lett. B 411, 9 (1997),
Erratum-ibid. Phys. Lett. B 421, 413 (1998).

\bibitem{kt98} K. Tsushima, \textit{et al.,} Phys. Rev. C 59, 369 (1999),
Erratum-ibid. Phys. Rev. C 61, 029903 (2000).

\bibitem{lzw} Z. Lin, C. M. Ko, and B. Zhang, Phys. Rev. C 61, 024904 (2000).

\bibitem{pdg14} K. A. Olive \textit{et al.} (Particle Data Group), Chin.
Phys. C, 38, 090001 (2014).

\bibitem{tb65} T. Bauer and D. R. Yennie, Phys. Lett. 60B, 165 (1976).

\bibitem{tb69} T. Bauer and D. R. Yennie, Phys. Lett. 60B, 169 (1976).

\bibitem{tb79} T. H. Bauer, \textit{et al.,} Rev. Mod. Phys. 50, 261 (1978);
51, 407(E) (1979).

\bibitem{ad87} A. Donnachie and P. V. Landshoff, Phys. Lett. B 185, 403
(1987).

\bibitem{ma96} M. A. Pichowsky and T. S. H. Lee, Phys. Lett. B 379, 1 (1996).

\bibitem{vp14} V. P. Goncalves and M. L. L. da Silva, Phys. Rev. D 89,
114005 (2014).

\bibitem{rj71} R. J. Eden, Rep. Prog. Phys. 34, 995 (1971).

\bibitem{wh02} W. H. Liang, P. N. Shen, J. X. Wang and B. S. Zou, J. Phys. G
28, 333 (2002).

\bibitem{rm87} R. Machleidt, K. Holinde, and Ch. Elster, Phys. Rep. 149, 1
(1987).

\bibitem{chu} https://www.jlab.org/Hall-C/talks/08\_21\_06/Chudakov.pdf.

\bibitem{al} A. Levy, arXiv:0711.0737.

\bibitem{E760} T. A. Armstrong \textit{et al.,} Phys. Rev. Lett. 69, 2337
(1992).

\bibitem{E835} M. Andreotti \textit{et al.,} Phys. Rev. D 72, 032001 (2005);
D. Joffe, arXiv:hep-ex/0505007.

\bibitem{HERMES} A. Movsisyan, EPJ Web Conf. 73 (2014) 02017 (2014)
\end{thebibliography}
\end{document}